\documentclass[11pt]{article}
\usepackage{amsmath}
\usepackage{amssymb}
\usepackage{theorem}
\usepackage{booktabs}
\usepackage{oldgerm}
\usepackage{epsfig}
\usepackage{graphicx}
\usepackage{epstopdf}
\usepackage{centernot}
\usepackage[algoruled,linesnumbered,noend]{algorithm2e}

\DeclareMathOperator{\Card}{Card}

\newtheorem{proposition}{Proposition}[section]

\newtheorem{definition}{Definition}[section]
\newtheorem{theorem}{Theorem}[section]
\newtheorem{conjecture}{Conjecture}[section]
\newtheorem{lemma}{Lemma}[section]
\newtheorem{remark}{Remark}[section]
\newtheorem{example}{Example}[section]
\newtheorem{corollary}{Corollary}[section]
\newcommand{\N}{\mathbb{N}}
\newcommand{\Z}{\mathbb{Z}}
\newcommand{\Zc}{\mathbb{Z}/n\mathbb{Z}}
\newcommand{\un}{\underline}
\renewcommand{\P}{\underline{P}}
\renewcommand{\S}{\underline{S}}
\newcommand{\X}{\underline{X}}
\newcommand{\A}{\underline{A}}

\def\petitcarre{\vrule height4pt width 4pt depth0pt}
\def\enddim{\relax\ifmmode\eqno{\hbox{\petitcarre}}
\else
{\unskip\nobreak\hfil\penalty50
   \hskip2em\hbox{}\nobreak\hfil
   \petitcarre
   \parfillskip=0pt \finalhyphendemerits=0
  \par\medskip}\fi}

\def \begdim {\noindent {\sc Proof} : \par \noindent}

\numberwithin{theorem}{section}
\numberwithin{equation}{section}
\numberwithin{figure}{section}
\numberwithin{table}{section}

\begin{document}

\date{}
\numberwithin{equation}{section}

\title{Finite maximal codes and factorizations of cyclic groups}

\author{Clelia De Felice
\thanks{
Dipartimento di Informatica,
Universit\`a degli Studi di Salerno,
via Giovanni Paolo II 132,
84084 Fisciano (SA), Italy.
$\qquad \qquad \qquad \qquad \qquad \qquad $
$\qquad \qquad \qquad$
$\qquad \qquad \qquad$
$\qquad \qquad \qquad$
$\qquad \qquad \qquad$
$\qquad \qquad \qquad$
{\tt E-mail: defelice@unisa.it}}}

\providecommand{\keywords}[1]{\textbf{\textit{Keywords:}} #1}

\thispagestyle{plain}

\maketitle

%%%%%%%%%%%%%%%%%%%%%%%%%%%%%%%%%%%%%%%%%%%%%%%%%%%%%%%%%%%%%

\begin{abstract}
Variable-length codes are the {\it bases} of the free
submonoids of a free monoid. There are some important longstanding open questions
about the structure of finite {\it maximal codes}, namely the
{\it factorization conjecture} and the
{\it triangle conjecture}, proposed
by Perrin and Sch\"{u}tzemberger. The latter
concerns finite codes $Y$ which are subsets of $a^* B a^*$,
where $a$ is a letter and $B$ is an alphabet not containing $a$.
A structural property of finite maximal codes has recently been shown by
Zhang and Shum. It exhibits a relationship between finite maximal codes
and factorizations of cyclic groups.
With the aim of highlighting the links between this result and other older ones on maximal and factorizing codes,
we give a simpler and a new proof of this result.
As a consequence, we prove that for any finite maximal code $X \subseteq (B \cup \{a \})^*$
containing the word $a^{pq}$, where $p,q$ are prime numbers,
$X \cap a^* B a^*$ satisfies the triangle conjecture.
Let $n$ be a positive integer that is a product
of at most two prime numbers.
We also prove that it is decidable whether a finite code
$Y \cup a^{n} \subseteq a^* B a^* \cup a^*$
is included in a finite maximal code and that, if this holds,
$Y \cup a^{n}$ is included in a code that also satisfies the factorization conjecture.
\end{abstract}

%%%%%%%%%%%%%%%%%%%%%%%%%%%%%%%%%%%%%%%%%%%%%%%%%%

%%%%%%%%%%%%%%%%%%%%%%%%%%%%%%%%%%%%%%%%%%%%%%%%%%%%%%%%%%%%%%%%%%

%%%%%%%%%%%%%%%%%%%%%%%%%%%%%%%%%%%%%%%%%%%%%%%%%%%%%%%%%%%%%%%%%%

\section{Introduction} \label{IN}

The theory of {\it variable-length codes} takes its origin
in the framework of the theory of information, since Shannon's early
works in the 1950's. An algebraic theory of codes was subsequently initiated
by Sch\"{u}tzenberger,
who proposed in \cite{SC55} the semigroup theory as a
mathematical setting for the
study of these objects.
In this context the theory of codes has been extensively
developed, showing strong relations with automata theory, combinatorics
on words, formal languages and the theory of semigroups (see \cite{BPR}
for a complete treatment of this topic).
In this paper
we follow this algebraic approach and codes
are defined as the {\it bases} of the free
submonoids of a free monoid.

We are interested in some important longstanding open questions
about the structure of finite {\it maximal codes}
(maximal objects in the class of codes for the order of set inclusion).
One of these conjectures asks whether
any finite maximal code $X$ is {\it (positively) factorizing} \cite{SC}, that is if
there always exist
finite subsets $P$, $S$ of $A^*$ such that
\begin{eqnarray} \label{EFC}
\underline{X} - 1 & = & \underline{P}(\underline{A} - 1)\underline{S}
\end{eqnarray}
(here $1$ is the empty word and \underline{X} denotes
the {\it characteristic polynomial}
of a finite language $X$, i.e., the formal sum of its elements).

The above conjecture
was formulated by
Sch\"{u}tzenberger but, as
far as we know, it does not appear
explicitly in any of his papers.
It was quoted as the
{\it factorization conjecture} in \cite{P1}
for the first time and then also reported
in \cite{BPR}.
The major contribution to this conjecture is due to
Reutenauer \cite{REU83,REU85}. In particular,
he proved that for any finite
maximal code $X$ over $A$, there exist polynomials
$P, S \in \Z \langle A \rangle$ such that Eq. (\ref{EFC}) holds, that is,
$\underline{X} - 1 = P(\underline{A} - 1)S$.
Other partial results concerning this conjecture may be found in
\cite{BO78,BO79,BO81,DF93,DF13,DFReu,RES,ZHG}. Some of them show
a relationship between \textit{factorizations of finite cyclic groups}
and factorizing codes. We recall that a
pair $(R,T)$ of subsets of $\N$
is a {\it factorization}
of $\Zc$ if
for each $z \in \{0, \ldots , n-1 \}$
there exists a unique pair $(r,t)$, with $r \in R$
and $t \in T$,
such
that $r+t = z \pmod{n}$. The simplest example is provided by the
{\it Krasner factorizations}, where for each $z$ one has $r + t = z$ \cite{KRR}.

The factorization conjecture is still open and weaker forms of it have been proposed and reported below.

Two words $x,y$ are {\it commutatively equivalent} if
the symbols of $y$ can be reordered to make $x$.
Two sets $X,Y$ are commutatively equivalent if there is a bijection $\phi$ from
$X$ onto $Y$ such that for every $x \in X$, $x$ and $\phi(x)$ are commutative equivalent.
A well known class of codes is that
of {\it prefix} codes, i.e., codes such that
none of their words is a left factor of another.
A code $X \subseteq A^*$ is {\it commutatively prefix} if there exists a prefix
code $Y \subseteq A^*$ which is commutatively equivalent to $X$.

It is conjectured that every finite maximal code is commutatively prefix.
This is the {\it commutative equivalence conjecture},
due Sch\"{u}tzenberger \cite{SC65} and inspired by a problem of information theory
\cite{PS77b}.
Any factorizing code is commutatively prefix.
Partial results on the commutative equivalence conjecture have been
proved in \cite{PS77b,MR}.

A third conjecture takes into account {\it bayonet} codes, i.e., codes such that each of their words has the form $a^iba^j$,
$a \in A$, $b \in A \setminus \{a \}$. It is conjectured that for any finite bayonet code $X$ which can be embedded
in a finite maximal code, one has
\begin{equation} \label{introduzione}
\Card(X) \leq \max\{|x| ~|~ x \in X \}
\end{equation}
This is the {\it triangle conjecture}, due to Perrin and Sch\"{u}tzemberger \cite{PS81}.
If $X$ is a finite maximal code and $X$ is commutatively prefix, then $X \cap a^*ba^*$
verifies the triangle conjecture, for any $a,b \in A$.
Partial results on the triangle conjecture have been proved in \cite{DF82,HAN,PS,ZHSH}.

Originally the three conjectures were proposed for codes with
no additional hypothesis. In 1985 Shor found a bayonet code $X$ such that
$\Card(X) > \max\{|x| ~|~ x \in X \}$ \cite{Shor}. Other counterexamples
may be found in \cite{Cordero19}. Thus the conjectures were restricted
as above to the smaller class of finite maximal codes and its subsets.

Notice that there are finite codes which are not contained in any finite maximal code
\cite{RES}. The {\it inclusion problem}, for a finite code $X$, is the existence of a finite maximal
code containing $X$. The {\it inclusion conjecture} claims that the inclusion problem
is decidable.

The aim of this paper is on the one hand to highlight the links between a recent result
on finite maximal codes and other less recent ones on maximal and factorizing codes,
on the other hand to deduce from these connections new results on the aforementioned conjectures.
More specifically, the
starting point of our research is a structural property of the maximal finite codes
which has recently been established by Zhang and Shum in \cite{ZHSH,ZHSHResGat}.
Let $X$ be a finite maximal code over $A$ such that $a^n \in X$, where $a \in A$.
Set $B = A \setminus \{ a \}$.
Zhang and Shum considered sets $X_w$ of words of the form $a^iwa^j$, where $w \in B(a^*B)^*$,
and $a^iwa^j \in X^* \setminus [a^n(a^*wa^* \cap X^*) \cup (X^* \cap a^*wa^*)a^n]$. They proved that the words in $X_w$
can be arranged in a matrix $(a^{r_{p,q}}wa^{v_{p,q}})_{1 \leq p \leq m,  1 \leq q \leq \ell}$
such that for each
$R_p = \{r_{p,q} ~|~ q \in \{1, \ldots , \ell \} \}$
and each
$T_q =\{v_{p,q} ~|~ p \in \{1, \ldots , m \} \}$, the pair
$(R_p,T_q)$ is a factorization of the
finite cyclic group of order $n$ (Theorem \ref{ZHmain}).
Our first result is a new simpler proof of the property established by
Zhang and Shum (Theorem \ref{endchinois}).
It is based on the so-called weak form of the factorization conjecture proved by Reutenauer
in \cite{REU85} and then by Zhang and Gu in \cite{ZG92} (Theorem \ref{WFc}).
Both in the proof of Zhang and Shum and in the one presented in this article,
a crucial role is played by the existence of a special ``universal'' factorization of a cyclic group associated with
all $X_w$, called here a {\it companion factorization} of $X$ (see Definition \ref{companion}).
We have proved that the case where this companion factorization is a Krasner factorization
is equivalent to a special arrangement of the words of $X_w$,
named a {\it good arrangement} in \cite{CDF05}
where it was introduced
for a factorizing code $X$ and $w \in B = A \setminus \{ a \}$
(Proposition \ref{krcompanionGood}).

In the same article \cite{ZHSH}, Zhang and Shum have proposed a stronger version of the triangle conjecture,
that is, they ask whether Eq. (\ref{introduzione}) holds for all $X_w$ when we change $w$ to $b$.
As a corollary of their result, they obtained that if $X$ has a Krasner pair as a companion factorization,
then $X$ satisfies this stronger version of the conjecture. Again, we give another proof of this result,
based on the properties of a good arrangement (Proposition \ref{triangleKrasner}).

It is known that if $X$ has a Krasner pair as a companion factorization, then
the factorizations $(R_p,T_q)$ are of a special type, named {\it Haj\'{o}s factorizations}
(Section \ref{HFRC}).
Not all factorizations of $\Zc$ are of this type but they certainly are for some values of $n$
which have been exactly identified \cite{Szabo}. The simplest case is
$n = p$ or $n = pq$, with $p$ and $q$ prime numbers.
We wondered if the condition that all $(R_p,T_q)$ are Haj\'{o}s factorizations guarantees
that the triangle conjecture (``basic'' or stronger version) is true for $X_w$.
This is an open problem and a main issue arising from this research.
However, Zhang and Shum proved that the answer is positive if $n = p$
and we have proved that the same is true for $n = pq$. That is, if $X$ is a finite maximal code and
$a^{pq} \in X$ with $p,q$ prime numbers, then $X \cap a^*Ba^*$ satisfies the triangle conjecture
(Corollary \ref{triangle2primes}).
Actually, for these codes we have proved more. Namely, let $n$ be a positive integer that is a product
of at most two prime numbers. In Proposition
\ref{inclusion}, we prove that a finite code
$Y \cup a^{n} \subseteq a^*Ba^* \cup a^*$
is included in a factorizing code if and only if it is included in a finite maximal code
and both conditions are equivalent to a decidable property for
$Y \cup a^{n}$. Consequently we state that it is decidable whether
$Y \cup a^{n}$ is included in a finite maximal code.

This paper is organized as follows.
Basics on words, codes and related notions are mainly collected in Section
\ref{basics}. Other more specific definitions and known results
are presented just before their use
in Sections \ref{FMC}, \ref{TrCj} and \ref{HFRC}.
Sections \ref{main} and \ref{TC}
are devoted to the proofs of our results.
Precisely, in Section \ref{main}
we present the new proof of the previously mentioned result of Zhang and Shum.
Then, we prove our new results in Section \ref{TC}.
Future research directions are discussed in
Section \ref{end}.
Finally, we have gathered in an appendix (Section \ref{App})
additional information on known results and definitions
that are quickly and informally mentioned in the article,
for the convenience of the reader who wants to give a deeper look to the problem.

%%%%%%%%%%%%%%%%%%%%%%%%%%%%%%%%%%%%%%%%%%%%%%%%%%%%%%%%%%%%%%%%%%%
\section{Basics} \label{basics}

%%%%%%%%%%%%%%%%%%%%%%%%%%%%%%%%%%%%%%%

\subsection{Words}

Let $A^{*}$ be the {\it free monoid}
generated by a finite alphabet $A$
and let $A^+=A^{*} \setminus 1$ where $1$ is
the empty word.
For a set $X$, $\Card(X)$ denotes the cardinality of $X$.
For a word $w \in A^*$, we denote by $|w|$ the {\it length}
of $w$.
The {\it reversal} of a word $w = a_1 \ldots a_n$,
$a_i \in A$, is the word $\widetilde{w} = a_n \ldots a_1$
and we set $\widetilde{X} = \{\widetilde{w} ~|~ w \in X \}$.
A word $x \in A^*$ is a {\it factor} of $w \in A^*$ if there are
$u_1,u_2 \in A^*$ such that $w=u_1xu_2$.
If $u_1 = 1$ (resp. $u_2 = 1$), then $x$ is a {\it prefix}
(resp. {\it suffix}) of $w$.
A factor (resp. prefix, suffix) $x$ of $w$
is {\it proper} if $x \not = w$.
A set $X \subseteq A^*$ is {\it rational} (or recognizable) if it is accepted by a
finite automaton.
Given $n$ sets $X_1, X_2, \ldots X_n \subseteq A^*$, with $n \geq 2$, the product
$X_1X_2 \cdots X_n$ is said to be {\it unambiguous} if any word $w \in X_1X_2 \cdots X_n$
has only one factorization $w = x_1x_2 \cdots x_n$ with $x_i \in X_i$, $1 \leq i \leq n$.
In this case, $\Card(X_1X_2 \cdots X_n) = \prod_{i = 1}^n \Card(X_i)$.

%%%%%%%%%%%%%%%%%%%%%%%%%%%%%%%%%%%%%%%%%%%%%%%%%%%%%%%%%
\subsection{Codes}

A subset $X$ of $A^{*}$ is a {\it code} over $A$ if
for all $h, k \geq 0$ and
$x_1, \ldots , x_h, x'_1, \ldots , x'_k \in X$, the relation
\begin{equation}
x_1x_2\cdots x_h =x'_1x'_2\cdots x'_k \notag
\end{equation}
implies
\begin{equation}
h = k \quad\mbox{ and }\quad x_i = x'_i \quad\mbox{for }\quad i = 1, \ldots , h\,. \notag
\end{equation}

The following is part of Propositions 2.2.3 and 2.2.5 in \cite{BPR}.

\begin{theorem} \label{stabile}
If $X \subseteq A^*$ is a code, then $X^*$ is stable, that is,
for all $u, v, x \in A^*$, we have
$$u, v, ux, xv \in X^* \quad \Rightarrow \quad x \in X^*$$
\end{theorem}

A set $X \subseteq A^+$ such that $X \cap XA^+ = \emptyset$
is a {\it prefix} code. $X$ is a {\it suffix} code if
$\widetilde{X}$ is a prefix code and $X$ is a {\it biprefix}
code when
$X$ is both a suffix and a prefix code.

The following is Proposition 2.1 in \cite{DFR85} (see also \cite{PS77,CCTesi}).

\begin{proposition} \label{nbaionette}
Let $A = \{a, b \}$ a two-letter alphabet.
If $X \cup \{ a^n \}$ is a code and $X \subseteq a^*ba^*$, then
$\Card(X) \leq n$.
\end{proposition}

%%%%%%%%%%%%%%%%%%%%%%%%%%%%%%%%%%%%%%%%%%%%%%%%%%%%%%%%%
\subsection{Maximal codes}

A code $X$ is a {\it maximal} code over $A$ if for each code $X'$
over $A$ such that $X \subseteq X'$ we have $X = X'$.
A code $X$ over $A$ is {\it complete} if for any $w \in A^*$,
one has $A^*wA^* \cap X^* \not = \emptyset$.
The two aforementioned notions are linked by the following fundamental theorem
\cite{BPR}.

\begin{theorem}
Any maximal code is complete. Any rational and complete code is maximal.
\end{theorem}

Then the class of finite maximal codes is closed under reversal, that is, if
$X$ is a finite maximal code, so is $\widetilde{X}$.
We recall that a word $w$ is {\it strongly right completable}
with respect to $X$
if, for all $u \in A^*$, there exists $v \in A^*$ such that $wuv \in X^*$
\cite{BPR}. If $X$ is a finite maximal code, then the set
of strongly right completable words with respect to $X$ is nonempty
\cite[Exercise 4, p. 75] {Salomaa}.
If $X$ is a finite maximal code, for each letter $a \in A$, there is an integer
$n \in \N$ such that $a^n \in X$, called the {\it order} of $a$ relative to $X$.

An interesting class of rational maximal prefix codes
named {\it semaphore codes} was introduced by Sch\"utzenberger
in \cite{Schutz64}. A set $X$ is a semaphore code if it has the form given in Eq. (\ref{eqSem}).
The following is Proposition 3.5.1 in \cite{BPR}.

\begin{proposition} \label{semaphore}
For any nonempty subset $S$ of $A^+$, the set
\begin{equation} \label{eqSem}
X = A^*S \setminus A^*SA^+
\end{equation}
is a maximal prefix code.
\end{proposition}

The following result is a direct consequence of Proposition \ref{semaphore}
(see also \cite[Example 3.5.2] {BPR}).

\begin{corollary} \label{semaphorew}
Let $A$ be an alphabet with at least two letters and let
$a \in A$. Let $B = A \setminus \{ a \}$.
For every $w \in B(a^*B)^*$,
the set $a^*w$ is a maximal prefix code.
\end{corollary}
\begdim
It suffices to apply Proposition \ref{semaphore} together with the observation
that $a^*w$ has the form given by Eq. (\ref{eqSem}) when $S = \{ w \}$.
\enddim

%%%%%%%%%%%%%%%%%%%%%%%%%%%%%%%%%%%%%%%%%%%%%%%%%%

\subsection{Composition} \label{Com}

Composition is a partially binary
operation on codes. We recall below the definition of
this operation and of the converse notion of decomposition.

Let $Z \subseteq A^*$, $Y \subseteq B^*$
be two codes such that $B = alph(Y)$
(i.e., each letter $b \in B$ is a factor of at least
one word in $Y$). Then the codes $Y$ and $Z$ are
{\it composable}
if there is a bijection from $B$ onto $Z$. If $\beta$ is
such a bijection, then $Y$ and $Z$ are called composable
{\it through} $\beta$. Then $\beta$ defines a morphism
from $B^*$ into $A^*$ which is injective since $Z$ is a code.
The set
$$X = \beta(Y) \subseteq Z^*$$
is a code over $A$.
We denote it by
$$X = Y \circ Z$$
and we say that $X$ is obtained by
{\it composition} of $Y$ and $Z$.
The words in $X$ are obtained just by replacing, in the words
of $Y$, each letter $b$ by the word $\beta(b) \in Z$.
The following result, proved in \cite[Proposition 2.2.6] {BPR},
underlines a second aspect of the composition operation, namely
the {\it decomposition} of a code into simpler ones.

\begin{proposition} \label{Dec1}
Let $X, Z \subseteq A^*$ be codes. There exists a
code $Y$ such that $X = Y \circ Z$ if and only if
$$X \subseteq Z^* \quad { and } \quad alph_Z(X) = Z$$
where the second condition above means that
all words in $Z$ appear in at least one factorization of a
word in $X$ as product of words in $Z$.
\end{proposition}

%%%%%%%%%%%%%%%%%%%%%%%%%%%%%%%%%%%%%%%%%%%%%%%%%%%%%%%%%%%%%%
\subsection{Polynomials}

Let $\Z \langle A \rangle$
(resp. $\N \langle A \rangle$)
denote the semiring of the {\it
polynomials} with noncommutative variables in $A$
and integer (resp. nonnegative integer)
coefficients.
For a finite subset
$X$ of $A^{*}$,  $\underline{X}$ denotes its
{\it characteristic polynomial}, defined by
$\underline{X}= \sum_{x \in X} x$.
Therefore, ``characteristic polynomial'' will be synonymous
with ``polynomial with coefficients $0,1$''.
The map which associates the polynomial $\sum_{n \in \N} (H,n)a^{n} \in \N[a]$
to a finite multiset $H$ of
nonnegative integers, is a bijection from
the set of the finite multisets $H$ of nonnegative integers
onto $\N[a]$.
We represent this bijection by the notation
$a^H = \sum_{n \in \N} (H,n)a^{n}$.
For example, $a^{\{0, 0, 1, 1, 1, 3, 4 \}} = 2 + 3a + a^3 + a^4$.
Therefore, if
$H_1, H_2, \ldots H_k \in \N \langle 1 \rangle$, the expression
$a^{H_1}ba^{H_2} \ldots a^{H_k}$ is
a notation for the product of
the formal power series
$a^{H_1},b,a^{H_2}, \ldots , a^{H_k}$.
For instance, $a^{\{2,3\}}ba^{\{1,5\}}=
a^2ba + a^2ba^5 + a^3ba + a^3ba^5$.
Computation rules are also defined:
$a^{M+L}=a^{M}a^{L}$,
$a^{M \cup L}= a^{M} + a^{L}$,
$a^{\emptyset}=0$, $a^0=1$.
Let $X, Y \subseteq \N$,
let $n \in \N$.
We write
$X = Y \pmod{n}$ if there is a bijection $\phi$ from
$X$ onto $Y$ such that for each $x \in X$,
$x = \phi(x) \pmod{n}$.
If such a bijection exists from $X$ onto a subset of
$\{0, \ldots , n-1 \}$, then this subset will be denoted by
$X_{(n)}$. If $x \in X$ and $n$ is clear from the context, we denote by $\overline{x}$
the residue of $x$ modulo $n$.

%%%%%%%%%%%%%%%%%%%%%%%%%%%%%%%%%%%%%%%%%%%%%%%%%%%%%%%%%%%%%%%%%%%%%%%
\subsection{Factorizations of cyclic groups} \label{FattG}

The notion of factorization of a finite abelian group was introduced for
the first time by Haj\'{o}s when he solved one of
Minkowski's conjectures by giving it a group-theoretical formulation
\cite{HA50a,SZSA}.
Let $G$ be a finite abelian group where the composition law
is written additively.
Given two subsets $R, T$ of $G$, we write
$R + T = \{r + t  ~|~ r \in R, t \in T \}$.
The sum $R + T$ is {\it direct} if for any element $g$ in $G$, there exists
at most one pair $(r, t)$ with $r \in R$ and $t \in T$ such that
$g = r + t$.
A sequence $S_{1},\ldots,S_{k}$
of subsets of $G$ is a {\it factorization} of $G$ (or $G$ is the
{\it direct sum} of its subsets $S_{i}$) if each element of $G$
may be written uniquely as a sum with just one term from each
$S_{i}$. Then $S_i$ is called a {\it factor} of $G$.

We shall be interested in factorizations
of cyclic groups. As usual,
we realize the cyclic group of order $n$
as the factor group $\Z/n\Z$ of the integers modulo
$n$. For the relation with codes, we will always consider
{\it positive} representatives of its classes.
Moreover, we will focus our attention on the factorizations of
$\Zc$ by two factors, as in the following definition.

\begin{definition}  \label{fattGC}
A pair $(R, T)$ of subsets of $\N$
is a factorization
of $\Zc$ if for any $z \in \{0, \ldots , n-1 \}$
there exists a unique pair $(r,t)$, with $r \in R$
and $t \in T$, such
that $r + t = z \pmod{n}$.
\end{definition}

Here, the classical hypotheses
$R,T \subseteq \{0, \ldots , n-1 \}$ and $0 \in R \cap T$
are not implicitly assumed to hold, as we consider factorizations of
$\Zc$ in relation to codes.
Note that if $(R, T)$ is a factorization of $\Zc$, then
$\Card(R + T) = \Card(R) \Card(T) = n$.
The following proposition is more or less classically known \cite{CDF96}.

\begin{proposition}
A pair $(R,T)$ of subsets of $\N$
is a factorization
of $\Zc$ if and only if there exists a finite
subset $H$ of $\N$ such that
\begin{equation}
a^{R}a^{T} = (\frac{a^{n}-1}{a-1})(1+a^{H}(a-1)) = \frac{a^{n}-1}{a-1} + a^H (a^n - 1) \notag
\end{equation}
\end{proposition}

The general structure of
the factorizations $(R,T)$ of $\Zc$ is still unknown but two simple
families of these pairs can be recursively constructed:
{\it Krasner factorizations} and {\it Haj\'{o}s factorizations}.
We recall that a Krasner factorization (of order $n$) is a pair $(I,J)$ of subsets of $\N$
such that for any $z \in \{0, \ldots , n-1 \}$ there exists a unique
$(i,j)$, with $i \in I$ and $j \in J$,
such that
$i+j=z$.
The pairs $(I,J)$ as above have been
completely described in \cite{KRR}.
Haj\'{o}s factorizations will be taken into account in Section \ref{HFRC}.

%%%%%%%%%%%%%%%%%%%%%%%%%%%%%%%%%%%%%%%%%%%%%%%%%%%%%%%%%%%%%%%%%%

\subsection{The factorization conjecture}

Given a finite maximal code $X$, a {\it factorization}
$(P,S)$ for $X$ is a pair of polynomials
$P,S \in \Z \langle A \rangle$
such that $\X=P(\A-1)S+1$.
Theorem \ref{ine} is from \cite{REU85}. It shows that any finite maximal code has a factorization.

\begin{theorem} \label{ine}
Let $X \in \N \langle A \rangle$,
with $(X,1)=0$, and let
$P,S \in \Z  \langle A \rangle$ be
such that $X=P(\A-1)S+1$. Then, $X$ is the characteristic polynomial
of a finite maximal code. Furthermore, if
$P,S \in \N \langle A \rangle$,
then $P,S$ are polynomials with coefficients $0,1$.
Conversely, for any finite maximal code $X$
there exist
$P,S \in \Z \langle A \rangle$
such that $\X=P(\A-1)S+1$.
\end{theorem}

Let $X$ be a finite maximal code.
A pair $(P,S)$ of subsets of $A^*$ is a
{\it positive factorization} for $X$ if
$(\P, \S)$ is a factorization for $X$.
Then $X$ is called a {\it (positively) factorizing code}.

Conjecture \ref{FatConj} is among the most difficult, unsolved
problems in the theory of codes. This
conjecture was formulated by
Sch\"{u}tzenberger but, as
far as we know, it does not appear
explicitly in any of his papers.
It was quoted as the
{\it factorization conjecture} in \cite{P1}
for the first time and then also reported
in \cite{BPR}.

\begin{conjecture} \label{FatConj} \cite{SC}
Any finite maximal code is a positively factorizing code.
\end{conjecture}

Finite maximal prefix codes are the simplest
examples of positively factorizing codes.
Indeed, $X$ is a finite maximal prefix code if and
only if $\X = \P(\A-1) + 1$ for a finite subset
$P$ of $A^*$ \cite{BPR}.
In the previous relation,
$P$ is the set of the proper prefixes
of the words in $X$.

As already said in Section \ref{IN}, the factorization conjecture is still open
and weaker forms of it have been proposed. In particular, the triangle conjecture will be recalled in Section \ref{TrCj}
and new results on it will be presented in Section \ref{TC}.

%%%%%%%%%%%%%%%%%%%%%%%%%%%%%%%%%%%%%%%%%%%%%%%%%%
\subsection{Further notations and assumptions}

Unless explicitly stated otherwise, from now on
$X$ will be a code over an alphabet $A$ containing at least two letters.
We assume that $X$ contains a power of a letter $a$, say $a^n$.
Then, we write $B = A \setminus \{ a \}$. Moreover, as in \cite{ZHSH},
for any $w \in B(a^*B)^*$, we set
\begin{eqnarray*}
X_w & = & (a^*wa^* \cap X^*) \setminus [a^n(a^*wa^* \cap X^*) \cup (a^*wa^* \cap X^*) a^n] \\
    & = & a^*wa^* \cap [ X^* \setminus (a^n X^* \cup X^* a^n)]
\end{eqnarray*}
Let $T$ be a finite set of nonnegative integers.
To simplify notation, the set $\{ a^t ~|~ t \in T \}$ will be denoted by
$a^T$.

%%%%%%%%%%%%%%%%%%%%%%%%%%%%%%%%%%%%%%%%%%%%%%%%%%%%%%%

\section{Some known results on finite maximal codes} \label{FMC}

In this section we focus on special sets $X_w$ of words previously defined.
If $X$ is a finite maximal code, then it is complete and the set $X_w$ is always nonempty (this is still
true if $w$ is any word in $A^*$, see \cite[Lemma 12.2.3]{BPR}
for a complete proof of this result).
Additional information about $X_w$ is provided through some propositions that follow.

\begin{proposition} \label{unica}
Let $X \subseteq A^*$ be a code on $A$ and let $a \in A$ be a letter such that $a^n \in X$.
For any $w \in B(a^*B)^*$ and any pair $(i,j)$ with $i, j \in \{0, \ldots , n-1 \}$
there exists at most one pair $(r , v)$, with
$r = i \! \pmod{n}$, $v = j \! \pmod{n}$ such that $a^rwa^v \in X_w$.
\end{proposition}
\begdim
It suffices to show that if $a^rwa^v, a^swa^t \in X_w$ with $r = s \! \pmod{n}$, $v = t \! \pmod{n}$,
then $r = s$ and $v = t$.
On the contrary, assume that $r = s + \lambda n$ with $\lambda > 0$.
One has $v \geq t$ or $v < t$.
If $v \geq t$, then $v = t + \mu n$, $\mu \geq 0$. Hence
$$a^rwa^v = (a^n)^{\lambda} a^swa^t (a^n)^{\mu} \in a^n (a^*wa^* \cap X^*)$$
in contradiction with $a^rwa^v \in X_w$.
Therefore, $t = v + \mu n$, with $\mu > 0$ and thus
\begin{eqnarray} \label{EQunica1}
a^rwa^v (a^n)^{\mu} & = & (a^n)^{\lambda} a^swa^t
\end{eqnarray}
Since $a^rwa^v, a^swa^t \in X_w$, there $x_1, \ldots , x_h, y_1, \ldots, y_k \in X$, $h,k \geq 1$,
such that
\begin{eqnarray} \label{EQunica2}
a^rwa^v & = & x_1 \cdots  x_h,  \quad a^swa^t = y_1 \cdots y_k
\end{eqnarray}
Moreover $x_1 \not = a^n$, otherwise $a^{r-n}wa^v = x_2 \cdots  x_h \in X^*$
and $a^rwa^v \in a^n (a^*wa^* \cap X^*)$.
By Eqs. \ref{EQunica1}, \ref{EQunica2}, we have
\begin{eqnarray*}
x_1 \cdots  x_h (a^n)^{\mu} & = & a^rwa^v (a^n)^{\mu}  =  (a^n)^{\lambda} a^swa^t = (a^n)^{\lambda} y_1 \cdots y_k \in X^*
\end{eqnarray*}
with $x_1 \not = a^n$ and $\lambda > 0$, contradicting the hypothesis that $X$ is a code.
Cases $r < s$ or $v \not = t$ may be handled in a similar way.
\enddim

Let $X \subseteq A^*$ be a finite maximal code on $A$. Let $n$ be the order of $a$,
$a \in A$. Following \cite{BPR}, for a word $w$ we denote by $C_a(w)$ the pairs of
residues modulo $n$ of integers $i, j \geq 0$ such that $a^iwa^j \in X^*$.
The following is Proposition 12.2.4 in \cite{BPR}.

\begin{proposition} \label{nelementiCA}
Let $X$ be a finite maximal code on the alphabet $A$. Let $a \in A$ be a letter
and let $n$ be the order of $a$. For each word $w \in A^*$, the set
$C_a(w)$ has exactly $n$ elements.
\end{proposition}

The following is a direct consequence of Proposition \ref{nelementiCA} and
was also obtained in \cite{ZHSH} as a corollary to Theorem \ref{ZHmain}.

\begin{proposition} \label{nelementi}
Let $X$ be a finite maximal code on the alphabet $A$. Let $a \in A$ be a letter
and let $n$ be the order of $a$. For each word $w \in B(a^*B)^*$, one has
$\Card(X_w) = n$.
\end{proposition}
\begdim
Let $X$ be a finite maximal code on the alphabet $A$. Let $a \in A$ be a letter
and let $n$ be the order of $a$. Let $w \in B(a^*B)^*$.
Let $\phi: X_w \rightarrow \{0, \ldots , n -1 \} \times \{0, \ldots , n -1 \}$
be the map defined by
$$\phi(a^rwa^v) = (\overline{r}, \overline{v})$$
If $a^rwa^v \in X_w$, then $(\overline{r}, \overline{v}) \in C_a(w)$, by definition of
$X_w$. Thus $\phi$ is a map of $X_w$ into $C_a(w)$.
We prove that $\phi$ is a bijection of $X_w$ onto $C_a(w)$.

If $y, z \in X_w$ are such that
$\phi(y) = \phi(z)$, then $y = a^rwa^v$, $z = a^swa^t$, with
$r = s \! \pmod{n}$, $v = t \! \pmod{n}$. Hence, by Proposition \ref{unica},
we have $y = a^rwa^v = a^swa^t = z$ and $\phi$ is injective.

Let $(i, j) \in C_a(w)$. By definition of $C_a(w)$, there is $(r , v)$, with
$r = i \! \pmod{n}$, $v = j \! \pmod{n}$ such that $a^rwa^v \in X^*$.
If we choose $r$ and $v$ minimal with respect to this condition, the word
$a^rwa^v$ turns out to be in $X_w$ and $\phi(a^rwa^v) = (i, j)$. Therefore
$\phi$ is surjective.

We proved that $\phi$ is a bijection of $X_w$ onto $C_a(w)$ hence,
by Proposition \ref{nelementiCA}, $\Card(X_w) = n$.
\enddim

Let $X$ be a code containing $a^n$.
Proposition \ref{Xwcodice} shows that the set $X_w \cup a^n$ is still a code.
Moreover, if $X$ is finite and maximal, then $X_w \cup \{a^n \}$ is a finite code.
The same statement has been proved in \cite[Theorem 2.3]{ZHSHResGat}
but for finite maximal codes $X$ and sets $X_w \cup a^n$ that verify the conditions
of Theorem \ref{ZHmain}, with a
proof that makes use of those conditions.
Theorem 2.3 in \cite{ZHSHResGat} proves more, namely that
$X_w \cup a^n$ is maximal in $a^*wa^* \cup a^*$.
The same additional result is stated with a different proof in
Proposition \ref{Xwcodicemax}.

\begin{proposition} \label{Xwcodice}
Let $X$ be a code on the alphabet $A$ and let $a \in A$ be a letter such that $a^n \in X$.
For each word $w \in B(a^*B)^*$, the set
$X_w \cup \{a^n \}$ is a code. If $X$ is a finite maximal code, then
$X_w \cup \{a^n \}$ is a finite code.
\end{proposition}
\begdim
Let $X$ be a code on the alphabet $A$ and let $a \in A$ be a letter such that $a^n \in X$.
Let $w \in B(a^*B)^*$.
The set $X_w \cup \{a^n \}$ is a code.
Indeed, suppose the contrary. Then there exists a word $z$ in $(X_w \cup \{a^n \})^+ \setminus a^*$,
of minimal length,
that has two distinct factorizations,
$$z = y_1 y_2 \cdots y_h = y'_1 y'_2 \cdots y'_k$$
($h, k \geq 1$, $y_i, y'_j \in X_w \cup a^n$).
Since $z \not \in a^*$, there exists minimal $h', k' \geq 1$ such that
\begin{equation} \label{EqCod1}
y_{h'} = a^rwa^s \in X_w, \quad y_{k'} = a^twa^m \in X_w
\end{equation}
Since $z$ is of minimal length, one between
$h'$ and $k'$ is necessarily equal to 1. Assume that $k' = 1$ (the case $h' = 1$ is symmetric),
thus we can rewrite Eq. (\ref{EqCod1}) as follows
\begin{equation}
y_1  =  y_2 = \ldots = y_{h'-1} = a^n, \quad y_{h'} = a^rwa^s, \quad y'_1 = a^twa^m \notag
\end{equation}
If $t < r + (h'-1)n$, then $a^tw$ is a proper prefix of $a^{r + (h'-1)n}w$,
in contradiction with Corollary \ref{semaphorew}.
Analogously, if $t > r + (h'-1)n$, then $a^{r + (h'-1)n}w$ is a proper prefix of
$a^tw$, which is impossible again by Corollary \ref{semaphorew}.
Hence $t = r + (h'-1)n$. Moreover, $m \not = s$ because $y'_1 = a^twa^m \in X_w$.
In conclusion, one of the following two cases holds
\begin{itemize}
\item[(1)]
$y_1  =  y_2 = \ldots = y_{h'-1} = a^n$,
$y_{h'} = a^rwa^s$, $y'_1 = a^twa^m = a^{r + (h'-1)n}wa^{s + \lambda}$, $\lambda > 0$
\item[(2)]
$y_1  =  y_2 = \ldots = y_{h'-1} = a^n$,
$y_{h'} = a^rwa^{m + \lambda}$, $y'_1 = a^twa^m = a^{r + (h'-1)n}wa^m$, $\lambda > 0$
\end{itemize}
Set $x = a^{\lambda}$.
Suppose that case (1) holds.
We notice that
\begin{eqnarray*}
u & = & y_1 \ldots y_{h'} \in a^{(h'-1)n}X_w \subseteq X^* \\
v & = & y'_2 \cdots y'_k \in (X_w \cup a^n)^* \subseteq X^* \\
ux & = & y_1 \ldots y_{h'}x = y'_1 \in X_w \subseteq X^*\\
xv & = & xy'_2 \cdots y'_k = y_{h'+1} \cdots y_h \in (X_w \cup a^n)^+ \subseteq X^*
\end{eqnarray*}
By Theorem \ref{stabile} applied to $X$, we have
$x = a^{\lambda} = a^{n \mu}$,
hence $y_{h'} = a^rwa^s$ and $y'_1 = a^{r + (h'-1)n}wa^{s + n\mu}$,
in contradiction with Proposition \ref{unica}.
Similar arguments apply if case (2) holds.
In this case, we set
\begin{eqnarray*}
u & = & y'_1 \in X_w \subseteq X^* \\
v & = &  y_{h'+1} \cdots y_h \in (X_w \cup a^n)^* \subseteq X^* \\
ux & = & y'_1a^{\lambda} = y_1 \ldots y_{h'} \in (X_w \cup a^n)^*\subseteq X^*\\
xv & = & xy_{h'+1} \cdots y_h = y'_2 \cdots y'_k \in (X_w \cup a^n)^+ \subseteq X^*
\end{eqnarray*}
The rest of the proof runs as before.

If $X$ is a finite maximal code, then $X_w \cup \{a^n \}$ is a finite set
by Proposition \ref{nelementi}.
\enddim

\begin{proposition} \label{Xwcodicemax}
Let $X$ be a finite maximal code on the alphabet $A$. Let $a \in A$ be a letter
and let $n$ be the order of $a$. For each word $w \in B(a^*B)^*$, the finite code
$X_w \cup \{a^n \}$ is maximal in $a^*wa^* \cup a^*$.
\end{proposition}
\begdim
Let $X$ be a finite maximal code on the alphabet $A$. Let $a \in A$ be a letter
and let $n$ be the order of $a$. Let $w \in B(a^*B)^*$. By Proposition \ref{nelementi},
$\Card(X_w) = n$.
If $X_w \cup \{a^n \}$ were not a maximal code in $a^*wa^* \cup a^*$,
then we could find $y = a^rwa^s \in a^*wa^* \setminus X_w$ such that $W = X_w \cup \{y \} \cup \{a^n \}$ would be still a code.
The set $Z = \{a, w \}$ is a prefix code.
Moreover $X_w \not = \emptyset$. Hence,
$W \subseteq Z^*$ and $alph_Z(W) = Z$, that is, all words in $Z$
appear in at least one factorization of a
word in $W$ as product of words in $Z$.
By Proposition \ref{Dec1}, there exists a code $Y$ over a two-letter alphabet $D = \{c, d \}$
such that $W$ is obtained by composition of $Y$ and $Z$ by means of a bijection $\beta$ from $D$ onto $Z$.
In particular $W = \beta(Y)$, where $\beta$ is now the induced injective morphism
from $D^*$ into $A^*$.
Set $\beta(c) = a$, $\beta(d) = w$.
Thus $c^n \in Y$ and
$\Card(Y) = \Card(\beta(Y)) = \Card(W) = n + 2 $
in contradiction with Proposition \ref{nbaionette}.
\enddim

Notice that some of these sets $X_w$
are subsets of $X$. For instance, $X_b \subseteq X$, for every $b \in B = A \setminus \{ a \}$.
Other examples are given below.
They are all codes on the two-letter alphabet $\{a, b \}$.

\begin{example} \label{EX1}
Consider the finite maximal code \cite[Example 4.3]{BDF92}
$$X = \{a^8, ba^2, ba^6, aba^2, aba^6, a^3b, a^2b, a^3ba^4, a^2ba^4, b^2, ab^2, bab, abab, b^2a^4, ab^2a^4, baba^4, ababa^4 \}$$
We have
\begin{eqnarray*}
X_b & = & \{ba^2, ba^6, aba^2, aba^6, a^3b, a^2b, a^3ba^4, a^2ba^4 \} \subseteq X \\
X_{bab} & = & \{bab, abab, baba^4, ababa^4, (a^3b)(aba^2), (a^3b)(aba^6), (a^2b)(aba^2), (a^2b)(aba^6) \}
\end{eqnarray*}
\end{example}

\begin{example} \label{EX2}
Consider the finite maximal code \cite[Example 5.1]{BDF92}
$$X = \{a^4, ab, a^3b, a^2ba, a^4ba, b^2, a^2b^2, ab^2a, baba, a^3b^2a, a^2baba, b^3a, a^2ba^3a \}$$
We have
\begin{eqnarray*}
X_b & = & \{ ab, a^3b, a^2ba, a^4ba \} \subseteq X \\
X_{bb} & = & \{ b^2, a^2b^2, ab^2a, a^3b^2a \} \subseteq X
\end{eqnarray*}
\end{example}

\begin{example} \label{EX3}
Consider the finite maximal code $X$ defined by the following positive factorization \cite[Example 3.1]{DF13}
\begin{eqnarray*}
\P & = & 1 + a^{2}ba^{\{0,1,2,3,4,5,6\}} + a^{2}ba^3ba^{\{0,1,2,3,4,5,6\}}, \\
\S & = & a^{\{0,1,2,3,4\}} + a^{\{ 0,1 \}}ba^{\{0,1,2,3,4\}}.
\end{eqnarray*}
That is,
\begin{eqnarray*}
\X - 1 & = & \P(\A -1)\S
\end{eqnarray*}
We have
\begin{eqnarray*}
&& \X_b  =  a^{2}ba^{\{7,8,9,10,11\}} \\
&& \X_{ba^2b}  =  a^{2}ba^2ba^{\{0,1,2,3,4\}} \\
&& \X_{ba^2bab}  =  a^{2}ba^2baba^{\{0,1,2,3,4\}} \\
&& \X_{ba^3ba^2bab}  =  a^{2}ba^3ba^2baba^{\{0,1,2,3,4\}}
\end{eqnarray*}
Notice that $X_b, X_{ba^2b}, X_{ba^2bab}, X_{ba^3ba^2bab}$ are all subsets of $X$.
More precisely, for any $w \in b(a^*b)^*$ which is a factor of $X$, we have $X_w \subseteq X$.
\end{example}

We deal with special arrangements of the words in a set $Y$ over a matrix.
Sometimes we will use the same symbol $Y$ to denote an arrangement of its words over a matrix
and even the same representation as a set, if the context does not make it ambiguous.
For short, we will refer to such an arrangement as an arrangement of $Y$.

Let
$\mathcal{X}_w = (a^{r_{k,m}}wa^{v_{k,m}})_{1 \leq k \leq s, \; 1 \leq m \leq t}$
be an arrangement of $X_w$.
Then $\mathcal{R} =
(r_{k,m})_{1 \leq k \leq s, \; 1 \leq m \leq t}$
is the {\it induced arrangement} of the rows
$R_{k} =\{r_{k,m} ~|~ m \in \{1, \ldots , t \} \}$
and $\mathcal{T} =
(v_{k,m})_{1 \leq k \leq s, \; 1 \leq m \leq t}$
is the {\it induced arrangement} of the columns
$T_{m} =\{v_{k,m} ~|~ k \in \{1, \ldots , s \} \}$.
Furthermore, $R_{k,w} =
\{a^{r_{k,m}}wa^{v_{k,m}} ~|~ 1 \leq m \leq t \}$
(resp. $T_{m,w} =
\{ a^{r_{k,m}}wa^{v_{k,m}} ~|~ 1 \leq k \leq s \}$)
is
a {\it word-row} (resp.
a {\it word-column})
of $X_w$, for $1 \leq k \leq s$
(resp. $1 \leq m \leq t$).

In Section \ref{RecallZS}, we state a structural property of sets $X_w$ which
has been proved in \cite{ZHSH,ZHSHResGat}. Then, in Section \ref{RecallReu}
we recall a result proved in \cite{REU85} that allows us to
establish the aforementioned property differently.

%%%%%%%%%%%%%%%%%%%%%%%%%%%%%%%%%%%%%%%%%%%%%%%%%%%%%%%%%%%%%%%%%%%%%%%%%%%%%%%%%%%%%%
\subsection{A structural property of finite maximal codes} \label{RecallZS}

Theorem \ref{ZHmain} is a part of \cite[Theorem 2.9] {ZHSH}.
It proves the existence of a special arrangement of the words of
$X_w$ in a matrix in which each pair formed by any row and any column
is a factorization of a cyclic group of order equal
to that of $a$.

\begin{theorem} \label{ZHmain}
Let $X \subseteq A^*$ be a finite maximal code and let $n$ be the order
of $a \in A$. There exists a pair of subsets of $a^*$,
$a^{P} = \{a^{p_1}, \ldots , a^{p_s} \}$ and
$a^{Q} = \{a^{q_1}, \ldots , a^{q_t} \} $
such that $(P,Q)$ is a factorization of $\Zc$ and, for any $w \in B(a^*B)^*$,
there exists an arrangement
$X_w = (a^{r_{k,m}}wa^{v_{k,m}})_{1 \leq k \leq s, \; 1 \leq m \leq t}$
of $X_w$ satisfying the following properties
\begin{itemize}
\item[(1)]
there exists
an ordered sequence $P_m=(p_{1,m}, \ldots , p_{s,m})$
of elements of $P$, $\quad 1 \leq m \leq t$, satisfying:
\begin{eqnarray} \label{EQZH1}
r_{1,m}+p_{1,m} &=& r_{2,m} + p_{2,m} = \ldots =
r_{s,m}+  p_{s,m} = q_m \pmod{n}.
\end{eqnarray}
\item[(2)]
there exists
an ordered sequence $Q_k=(q_{k,1}, \ldots , q_{k,t})$
of elements of $Q$, $\quad 1 \leq k \leq s$, satisfying:
\begin{eqnarray} \label{EQZH2}
v_{k,1} + q_{k,1} &=& v_{k,2} + q_{k,2} = \ldots =
v_{k,t} + q_{k,t} = p_k \pmod{n}.
\end{eqnarray}
\item[(3)]
For each row $R_k = \{ r_{k,m} ~|~ 1 \leq m \leq t \}$
and each column $T_m = \{v_{k,m}  ~|~ 1 \leq k \leq s \}$,
the pairs
$(R_k, T_m), (R_k, P), (Q, T_m)$ are factorizations of $\Zc$.
\end{itemize}
\end{theorem}

\begin{example} \label{EX1C}
Consider again the finite maximal code in Example \ref{EX1}, also reported below
$$X = \{a^8, ba^2, ba^6, aba^2, aba^6, a^3b, a^2b, a^3ba^4, a^2ba^4, b^2, ab^2, bab, abab, b^2a^4, ab^2a^4, baba^4, ababa^4 \}$$
Let $P = \{0, 4 \}$ and $Q = \{0, 1, 2, 3 \}$.
The pair $(P,Q)$ is a (Krasner) factorization of $\Z/8\Z$. An arrangement of $X_b$ as in Theorem \ref{ZHmain}
is the following
\begin{equation*}
X_b =
\begin{pmatrix}
ba^2 & aba^2 & a^2b & a^{3}b \\
ba^6 & aba^6 & a^2ba^4 & a^3ba^{4}
\end{pmatrix}
\end{equation*}
Similarly, an arrangement of $X_{bab}$ as in Theorem \ref{ZHmain}
is given below.
\begin{equation*}
X_{bab} = \begin{pmatrix}
bab & abab & a^2baba^2 & a^{3}baba^2 \\
baba^4 & ababa^4 & a^2baba^6 & a^3baba^{6}
\end{pmatrix}
\end{equation*}
\end{example}

\begin{example} \label{EX2C}
Let $X_b$ be as in Example \ref{EX2} and let $(P,Q)$ be the (Krasner) factorization of $\Z/4\Z$
with $P = \{0 \}$ and $Q = \{0, 1, 2, 3 \}$.
An arrangement of $X_b$ as in Theorem \ref{ZHmain}
is the following
\begin{equation*}
X_b =
\begin{pmatrix}
ab & a^2ba & a^3b & a^4ba
\end{pmatrix}
\end{equation*}
Similarly, an arrangement of $X_{bb}$ as in Theorem \ref{ZHmain}
is given below.
\begin{equation*}
X_{bb} = \begin{pmatrix}
bb & abba & a^2bb & a^{3}bba
\end{pmatrix}
\end{equation*}
\end{example}

\begin{example} \label{EX3C}
Let $X_b$ be as in Example \ref{EX3} and let $(P,Q)$ be the (Krasner) factorization of $\Z/5\Z$
with $P = \{0, 1, 2, 3, 4 \}$ and $Q = \{0 \}$.
An arrangement of $X_b$ as in Theorem \ref{ZHmain}
is the following
\begin{equation*}
X_b =
\begin{pmatrix}
a^2ba^7 \\
a^2ba^8 \\
a^2ba^9 \\
a^2ba^{10} \\
a^2ba^{11}
\end{pmatrix}
\end{equation*}
Arrangements for $X_{ba^2b}$, $X_{ba^2bab}$ and $X_{ba^3ba^2bab}$
as in Theorem \ref{ZHmain} are similar.
\end{example}

%%%%%%%%%%%%%%%%%%%%%%%%%%%%%%%%%%%%%%%%%%%%%%%%%%%%%%%%%%%%%%%%%%%%%%%%%%%%%%%%%%%%%%%%%%%
\subsection{A partial known result on the factorization conjecture} \label{RecallReu}

The following is \cite[Lemma 14.4.2 (ii)]{BPR}. It is a weak form of
the factorization conjecture proved by C. Reutenauer in
\cite{REU85}. It is also used in the proof of Theorem \ref{ine}.

\begin{theorem} \label{WFc}
For any finite maximal code $X$ there exist
finite subsets $P, S, P_1, S_1$ of $A^*$, with $1 \in P_1, S_1$ and
finite subsets $L_1, R_1$ of $A^+$ such that
$$\un{A^*} = \un{L_1} + \un{S} ~ \un{X^*} ~ \un{P_1} = \un{R_1} + \un{S_1} ~ \un{X^*} ~ \un{P} $$
\end{theorem}

\begin{remark} \label{vuota}
Let $X$ be a finite maximal code.
The proof of above theorem is based on some preliminary results.
In particular, in \cite[Lemma 14.4.2]{BPR} it has been proved the existence
of some special words $u_1, \ldots , u_d$, $v_1, \ldots , v_d$ with $u_1, v_1 \in X^*$, such that,
for any $1 \leq i, j \leq d$,
$$ \un{A^*} = \sum_{1 \leq \ell \leq d} u_i^{-1} \un{X^*} v_{\ell}^{-1} =
\sum_{1 \leq k \leq d} u_k^{-1} \un{X^*} v_{j}^{-1} $$
($d$ is a parameter associated with $X$, namely its degree).
In turn,
polynomials $P, S, P_1, S_1$ as in Theorem \ref{WFc}
are defined starting with these special words.
Notice that any $u_i$, $1 \leq i \leq d$, is a strongly right completable
word for $X$. Indeed by the above relation,
for all $u \in A^*$,  there exists $v_{\ell} \in A^*$,
$1 \leq \ell \leq d$, such that $u_iuv_{\ell} \in X^*$.
Analogously, any $v_j$, $1 \leq j \leq d$, is a strongly right completable
word for the reversal $\widetilde{X}$ of $X$.
The proof of Lemma 14.4.2 in \cite{BPR} gives an algorithm to compute these strongly
completable words.
\end{remark}

The proof of the first part of Proposition \ref{fattcompagna} is in
\cite[p. 443]{BPR}, it is reported here for the sake of completeness.

\begin{proposition} \label{fattcompagna}
Let $X \subseteq A^*$ be a finite maximal code and let $n$ be the order
of $a \in A$.
Let $S, P_1, L_1$ be as in Theorem \ref{WFc}.
Set $\un{P_a} = \un{P_1 \cap a^*} = a^R$,
$\un{S_a} = \un{S \cap a^*} = a^T$,
where $R, T \subseteq \N$.
Then the pair $(R,T)$ is a factorization of $\Zc$.
Moreover, if $X$ is a factorizing code, then $L_1 = \emptyset$ and
$(R,T)$ is a Krasner factorization of $\Zc$.
\end{proposition}
\begdim
Let $X$, $n$, $S, P_1, L_1$, $S_a$, $R$, $P_a$, $T$ be as in the statement.
By Theorem \ref{WFc}, we have
$$\un{A^*} = \un{L_1} + \un{S} ~ \un{X^*} ~ \un{P_1}$$
Set $a^W = \un{L_1 \cap a^*}$.
Taking $b = 0$ for all letters $b \not = a$, we obtain
$$a^* = a^T(a^n)^*a^R + a^W$$
Multiplying both sides by $1 - a^n$, we obtain
$$1 + a + \cdots + a^{n-1} = a^Ta^R + a^W(1 - a^n)$$
that is
$$1 + a + \cdots + a^{n-1} + a^W(a^n - 1) = a^Ta^R$$
hence $(R,T)$ is a factorization of $\Zc$.
In addition, if $X$ is a factorizing code and $(P_1, S)$ is a positive factorization for $X$, then
$L_1$ is empty. Therefore $W = \emptyset$, that is,
$a^W = 0$ and $(R,T)$ is a Krasner factorization of $\Zc$.
\enddim

In Section \ref{main} we will prove Theorem \ref{ZHmain}
by replacing the pair $(P, Q)$ with the pair $(R, T)$ defined
in Proposition \ref{fattcompagna}.

%%%%%%%%%%%%%%%%%%%%%%%%%%%%%%%%%%%%%%%%%%%%%%%%%%%%%%%%%%%
\section{A new proof}  \label{main}

The aim of this section is to give a new proof of Theorem \ref{ZHmain}.
It will be divided into two parts, proved in Proposition \ref{matrix} and
Theorem \ref{endchinois} respectively.
The notion of {\it companion factorization} defined below intervenes
in both statements.

\begin{definition} \label{companion}
Let $X$ be a code with $a^n \in X$. Let
$w \in B(a^*B)^*$ such that $X_w$ is nonempty.
A factorization $(T,R)$ of $\Zc$ is a
{\rm companion factorization} of $X_w$
if the product $a^T X_w a^R$ is unambiguous and one has
\begin{equation} \label{EqFatt1}
\{ a^{\overline{\ell_h}} w  a^{\overline{\ell_m}}  ~|~  a^{\ell_h} w  a^{\ell_m} \in a^T X_w a^R \}
= \{ a^h w a^m ~|~ h,m \in \{0, \ldots , n-1 \} \}
\end{equation}
A factorization $(T,R)$ of $\Zc$ is a
{\rm companion factorization} of $X$
if it is a companion factorization of any nonempty $X_w$,
for all $w \in B(a^*B)^*$.
\end{definition}

Proposition \ref{matrix} is
a consequence of Propositions \ref{nelementi} and \ref{fattcompagna}.

\begin{proposition} \label{matrix}
For any finite maximal code $X$ with $a^n \in X$,
there is a factorization $(T,R)$ of $\Zc$ that is a companion factorization of $X$.
\end{proposition}
\begdim
By Theorem \ref{WFc}, there are
finite subsets $S, P_1, L_1$ of $A^*$ such that
\begin{eqnarray} \label{EqWFc}
\un{A^*} & = & \un{L_1} + \un{S} ~ \un{X^*} ~ \un{P_1}
\end{eqnarray}
Set $P_a = P_1 \cap a^* = a^R$,
$S_a = S \cap a^* = a^T$,
where $R, T \subseteq \N$.
Let $w \in B(a^*B)^*$.
Set $Y_w = \{ a^h w a^m ~|~ h,m \in \{0, \ldots , n-1 \} \}$ and
let $f : a^T X_w a^R \rightarrow Y_w$ be the function defined
by $f(a^{\ell_h} w  a^{\ell_m}) = a^{\overline{\ell_h}} w  a^{\overline{\ell_m}}$.
We claim that $f$ is surjective. Indeed,
let $\lambda \in \N$ be such that
$\lambda n > \max\{ |y| ~|~ y \in L_1 \cup S \cup P_1 \}$.
By the definition of $\lambda$,
for any $h, m \in \{0, 1, \ldots , n-1 \}$ one has
$$(\un{A^*}, a^{\lambda n + h} w a^{\lambda n + m}) = 1, \quad (\un{L_1}, a^{\lambda n + h} w a^{\lambda n + m}) = 0$$
hence, by Eq. (\ref{EqWFc}), there is one and only one triple $(s, x, p)$ with $s \in S$, $x \in X^*$, $p \in P_1$.
\begin{eqnarray} \label{EqFatt2}
a^{\lambda n + h} w a^{\lambda n + m} & = &  s x p
\end{eqnarray}
Looking at the definition of $\lambda$ once again, we can refine this statement as follows:
for any $h, m \in \{0, 1, \ldots , n-1 \}$
there is one and only one triple $(a^t, a^{n \mu_1}a^{r'}wa^{t'}a^{n \mu_2}, a^r)$, with
$s = a^t \in a^T$, $x  = a^{n \mu_1}a^{r'}wa^{t'}a^{n \mu_2}$,
$a^{r'}wa^{t'} \in X_w$, $p = a^r \in a^R$, such that
\begin{eqnarray} \label{EqFatt3}
a^{\lambda n + h} w a^{\lambda n + m} & = & a^t a^{n \mu_1}a^{r'}wa^{t'}a^{n \mu_2}a^r
\end{eqnarray}
As a consequence of the above equation,
for each pair $(h, m)$, with $h,m \in \{0, \ldots , n-1 \}$,
there exists at least a pair $(\ell_h , \ell_m)$ of nonnegative integers, with
$\ell_h = h \! \pmod{n}$, $\ell_m = m \! \pmod{n}$,
such that
\begin{equation} \label{EqFatt4}
t + r' = \ell_h, \quad r + t' = \ell_m
\end{equation}
and moreover $a^t \in a^T$, $a^{r'}wa^{t'} \in X_w$, $a^r \in a^R$.
Therefore $f$ is surjective.
On the other hand, the product $a^T X_w a^R$ is unambiguous because it is a subset of
$S X^*P_1$. Furthermore, by Proposition \ref{nelementi}
one has $\Card(X_w) = n$ and by Proposition \ref{fattcompagna} one has
$\Card(a^T a^R) = n$. These two equalities yield
\begin{equation} \label{EqFatt6}
\Card(a^T X_w a^R) = \Card(T)\Card(X_w)\Card(R) = n^2 = \Card(Y_w)
\end{equation}
Hence, $f$ is bijective
and the proof is complete.
\enddim

The rest of the proof of Theorem \ref{ZHmain} is reported below.

\begin{theorem} \label{endchinois}
Let $X$ be a finite maximal code with $a^n \in X$, let
$(T,R)$ be a factorization of $\Zc$. If $(T,R)$ is a companion
factorization of $X$, then,
for any $w \in B(a^*B)^*$,
there is a bijection from $a^Rwa^T$
onto $X_w$. Moreover,
there is an arrangement
$\mathcal{X}_w = (a^{r_{p,q}}wa^{v_{p,q}})_{1 \leq p \leq m, \; 1 \leq q \leq \ell}$
of $X_w$
such that
\begin{itemize}
\item[(1)]
for each word-column
$T_{q,w} =
\{a^{r_{p,q}}wa^{v_{p,q}} ~|~ 1 \leq p \leq m \}$
of $\mathcal{X}_w$, $1 \leq q \leq \ell$
there exists
an ordered sequence $\mathcal{T}_q = (t_{1, q}, \ldots , t_{m, q})$
of elements of $T$, $1 \leq q \leq \ell$, satisfying:
\begin{eqnarray} \label{EQG1}
r_{1,q} + t_{1,q} &=& r_{2,q} + t_{2, q} = \ldots =
r_{m, q}+  t_{m , q} = r_q \pmod{n}
\end{eqnarray}
\item[(2)]
for each word-row
$R_{p,w} =
\{a^{r_{p,q}}wa^{v_{p,q}} ~|~ 1 \leq q \leq \ell \}$
of $\mathcal{X}_w$, $1 \leq p \leq m $
there exists
an ordered sequence $\mathcal{R}_p = (r'_{p, 1}, \ldots , r'_{p, \ell})$
of elements of $R$, $1 \leq p \leq m$, satisfying:
\begin{eqnarray} \label{EQG2}
v_{p,1} + r'_{p, 1} &=& v_{p, 2} + r'_{p, 2} = \ldots =
v_{p, \ell} +  r'_{p, \ell} = t_p \pmod{n}
\end{eqnarray}
\item[(3)]
for each row $R_p = \{r_{p,q} ~|~ q \in \{1, \ldots , \ell \} \}$
and and each column
$T_q =\{v_{p,q} ~|~ p \in \{1, \ldots , m \} \}$, the pairs
$(T , R_p)$, $(R_p,T_q)$, $(T_q, R)$ are all factorizations of
$\Zc$.
\end{itemize}
Finally, let $w, w' \in B(a^*B)^*$, let
$(R_{p,(w)},T_{q,(w)})$ $(R_{p',(w')},T_{q',(w')})$
be factorizations of $\Zc$
associated with $X_w$ and $X_{w'}$ respectively, as in item (3).
Then, $(R_{p,(w)},T_{q',(w')}), (R_{p',(w')}, T_{q,(w)})$ are factorizations of $\Zc$
and they are all companion factorizations of $X$.
\end{theorem}
\begdim
Let $X$ be a finite maximal code with $a^n \in X$.
Let $(T,R)$ be a factorization of $\Zc$ that
is a companion factorization of $X$. Let $w \in B(a^*B)^*$.
Set
\begin{eqnarray*}
R & = & \{r_1, \ldots , r_{\ell} \}, \\
T & = & \{t_1, \ldots, t_{m} \}
\end{eqnarray*}
Consider the matrix $\mathcal{D}$ associated with $a^Rwa^T$ and defined as follows
\begin{eqnarray} \label{A1}
\mathcal{D} & = & \begin{pmatrix}
              a^{r_1}wa^{t_1} & a^{r_2}wa^{t_1} & \ldots  & a^{r_{\ell}}wa^{t_1} \\
               a^{r_1}wa^{t_2} & a^{r_2}wa^{t_2} & \ldots  & a^{r_{\ell}}wa^{t_2} \\
                \quad         & \quad          & \ddots  & \quad \\
               a^{r_1}wa^{t_m} & a^{r_2}wa^{t_m} & \ldots  & a^{r_{\ell}}wa^{t_m} \\
               \end{pmatrix}
\end{eqnarray}
All rows are equal and each of them is equal to
$\{r_1, \ldots , r_{\ell} \} = R$.
Analogously, all columns are equal to $T$.
Thus $\mathcal{D}$ is an arrangement of $a^Rwa^T$ such that
for each row $R_p = \{r_{q} ~|~ q \in \{1, \ldots , \ell \} \}$
and each column
$T_q =\{t_{p} ~|~ p \in \{1, \ldots , m \} \}$,
$(T , R_p)$, $(R_p,T_q)$, $(T_q, R)$ are factorizations of
$\Zc$.
We now define an arrangement of $X_w$ which maintains the same property.

As a first step, we define a bijection from $a^Rwa^T$
onto $X_w$.
Let $(r_q, t_p) \in R \times T$. By Definition \ref{companion},
there are $t' \in T$, $a^iwa^j \in X_w$
and $r' \in R$ such that
\begin{equation} \label{map1}
r_q  =  i + t' \pmod{n}, \quad t_p  =  r' + j \pmod{n},
\end{equation}
that is,
\begin{equation} \label{map2}
r_q - t' = i \pmod{n}, \quad t_p - r'  =  j \pmod{n}
\end{equation}
Therefore, we set
\begin{equation} \label{map3bis}
\phi(a^{r_q}wa^{t_p})  =  a^iwa^j
\end{equation}
Then, $\phi$ is a map from $a^Rwa^T$ to $X_w$ because,
given $(r_q, t_p)$, the pairs $(t', i)$ and $(j, r')$ as in Eq. (\ref{map1}) are
uniquely determined by Definition \ref{companion}.
Hence a unique word in $X_w$ is associated with $a^{r_q}wa^{t_p}$ by $\phi$, namely
$a^iwa^j$ in Eq. (\ref{map3bis}), where $i,j$ are defined as in Eq. (\ref{map2}).

Moreover, $\phi$ is clearly injective. Indeed,
let $r_q, r_{q'} \in R$ and let $t_p, t_{p'} \in T$ be such that
$a^{r_q}wa^{t_p} \not = a^{r_{q'}}wa^{t_{p'}}$.
Thus, suppose that $r_q \not = r_{q'}$
(a symmetric argument applies if $t_p \not = t_{p'}$).
Let
$t'_1, i_1$ be associated with $r_q$ and let
$t'_2, i_2$ be associated with ${r_{q'}}$ by Eq. (\ref{map1}).
Then, by definition of $\phi$, one has
$\phi(a^{r_q}wa^{t_p}) = a^{i_1}wa^{j_1}$,
$\phi(a^{r_{q'}}wa^{t_{p'}}) = a^{i_2}wa^{j_2}$.
If we had $i_1 = i_2 \pmod{n}$, then by Eq. (\ref{map2}), we would also have
$r_q - t'_1 = r_{q'} - t'_2 \pmod{n}$, hence
$r_q + t'_2 = r_{q'} + t'_1 \pmod{n}$, which is impossible because by hypothesis $(R,T)$
is a factorization of $\Zc$ but $r_q \not = r_{q'}$.
Thus, $i_1 \not = i_2 \pmod{n}$ and $\phi(a^{r_q}wa^{t_p}) \not = \phi(a^{r_{q'}}wa^{t_{p'}})$.

Since $\phi$ is an injective map between two sets having the same cardinality, namely $n$,
$\phi$ is a bijection from $a^Rwa^T$ onto $X_w$.
We denote by $\mathcal{X}_w$ the arrangement of $X_w$ induced by $\mathcal{D}$ and $\phi$, that is,
$\mathcal{X}_w$ is obtained by replacing each element $a^{r_q}wa^{t_p}$ in $\mathcal{D}$ with
$\phi(a^{r_q}wa^{t_p})$.
Furthermore, to make the remainder of the proof clearer,
we slightly modify the notation in Eq. (\ref{map3bis}), namely
\begin{equation} \label{map3}
\phi(a^{r_q}wa^{t_p})  =  a^iwa^j = a^{r_{p,q}}wa^{v_{p,q}}
\end{equation}

As a second step, we notice that each word-column
$T_{q,w} =
\{a^{r_{p,q}}wa^{v_{p,q}} ~|~ 1 \leq p \leq m \}$
of $\mathcal{X}_w$, $1 \leq q \leq \ell$, is obtained by
applying $\phi$ to the corresponding word-column
$\{a^{r_q}wa^{t_{p}} ~|~ p \in \{1, \ldots , m \} \}$.
Looking at the definition of $\phi$, it is clear that
there exists
an ordered sequence $\mathcal{T}_q = (t_{1, q}, \ldots , t_{m, q})$
of elements of $T$, $1 \leq q \leq \ell$, satisfying Eq. (\ref{EQG1}):
\begin{eqnarray*}
r_{1,q} + t_{1,q} &=& r_{2,q} + t_{2, q} = \ldots =
r_{m, q}+  t_{m , q} = r_q \pmod{n}
\end{eqnarray*}
Similar arguments apply to the word-rows of $\mathcal{X}_w$ and allow us
to state that there exists
an ordered sequence $\mathcal{R}_p = (r'_{p, 1}, \ldots , r'_{p, \ell})$
of elements of $R$, $1 \leq p \leq m$, satisfying Eq. (\ref{EQG2}):
\begin{eqnarray*}
v_{p,1} + r'_{p, 1} &=& v_{p, 2} + r'_{p, 2} = \ldots =
v_{p, \ell} +  r'_{p, \ell} = t_p \pmod{n}
\end{eqnarray*}

Finally, we notice that
each row $R_p = \{r_{p,q} ~|~ q \in \{1, \ldots , \ell \} \}$ in $\mathcal{X}_w$
has $\ell = \Card(R)$ elements
and each column
$T_q =\{v_{p,q} ~|~ p \in \{1, \ldots , m \} \}$ has $m = \Card(T)$ elements,
with $\ell m = \Card(R) \Card(T) = n$. Thus,
to prove that $(T , R_p)$, $(R_p,T_q)$, $(T_q, R)$ are all factorizations of
$\Zc$, it is sufficient to show that for each of these pairs the sum is direct.

The sum $T + R_p$ is direct.
By contradiction, assume that there are $r_{p,q}, r_{p,q'}$ in $R_p$
and $t'_{p,q}, t'_{p,q'}$ in $T$, $1 \leq q, q' \leq \ell$, such that
\begin{equation} \label{EQG3}
r_{p,q} + t'_{p,q} = r_{p,q'} + t'_{p,q'} \pmod{n}, \quad r_{p,q} \not = r_{p,q'}
\end{equation}
By Eq. (\ref{EQG2}), there are $r'_{p,q}, r'_{p,q'}$ in $R$ such that
\begin{equation} \label{EQG4}
v_{p,q} + r'_{p, q} = v_{p,q'} + r'_{p, q'} = t_{p} \pmod{n}
\end{equation}
Moreover, $a^{r_{p,q}}wa^{v_{p,q}}$ and $a^{r_{p,q'}}wa^{v_{p,q'}}$ are in $X_w$, thus
$a^{t'_{p,q}}a^{r_{p,q}}wa^{v_{p,q}}a^{r'_{p, q}}$ and
$a^{t'_{p,q'}}a^{r_{p,q'}}wa^{v_{p,q'}}a^{r'_{p, q'}}$ are in $a^TX_wa^R$.
By Eqs. (\ref{EQG3}), (\ref{EQG4}), this is in contradiction with Definition \ref{companion}.
A similar argument applies for the columns, showing that the sum
$T_q + R$ is direct.

We now prove that the sum $R_p + T_q$ is direct.
By contradiction, assume that there are $r_{p,q_1}, r_{p,q_2}$ in $R_p$
and $v_{p_1,q}, v_{p_2, q}$ in $T_q$, such that
\begin{eqnarray*}
r_{p,q_1} &\not =& r_{p,q_2} \mbox{ or }  v_{p_1,q} \not = v_{p_2, q} \\
v_{p_1,q} + r_{p,q_1} & = & v_{p_2, q} + r_{p, q_2} \pmod{n}
\end{eqnarray*}
Moreover, we may assume that there are $\alpha \in \N$ such that one has
\begin{equation} \label{EQG5}
v_{p_1,q} + r_{p,q_1}  =  v_{p_2, q} + r_{p, q_2} + \alpha n
\end{equation}
As in the previous cases, $a^{r_{p,q_1}}wa^{v_{p,q_1}}$, $a^{r_{p,q_2}}wa^{v_{p,q_2}}$,
$a^{r_{p_1,q}}wa^{v_{p_1,q}}$ and $a^{r_{p_2,q}}wa^{v_{p_2,q}}$ are all elements
in $X_w \subseteq X^*$.
Moreover, by Eqs. (\ref{EQG1}), (\ref{EQG2}), there are $t_{p_1, q}, t_{p_2, q}$
in $T$ and $r'_{p,q_1}, r'_{p,q_2}$ in $R$ such that
\begin{eqnarray} \label{EQG6}
r_{p_1,q} + t_{p_1,q} &=&
r_{p_2, q} +  t_{p_2 , q} = r_q \pmod{n}, \\ \label{EQG7}
v_{p,q_1} + r'_{p, q_1} & = & v_{p,q_2} + r'_{p, q_2} = t_{p} \pmod{n}
\end{eqnarray}
Set $z = wa^{v_{p_1,q}}a^{r_{p,q_1}}w$.
The word
$$a^{r_{p_1,q}}za^{v_{p,q_1}} = a^{r_{p_1,q}}wa^{v_{p_1,q}}a^{r_{p,q_1}}wa^{v_{p,q_1}}$$
is in $X_z = a^*za^* \cap [ X^* \setminus (a^n X^* \cup X^* a^n)]$ because
$a^{r_{p_1,q}}wa^{v_{p_1,q}}$, $a^{r_{p,q_1}}wa^{v_{p,q_1}}$ are both $X_w$ and $X$ is a code containing $a^n$.
By similar arguments we can see that
$$a^{r_{p_2,q}}za^{v_{p, q_2}} = a^{r_{p_2,q}}wa^{v_{p_2,q}} a^{\alpha n} a^{r_{p,q_2}}wa^{v_{p,q_2}}$$
is in $X_z$. By Eqs. (\ref{EQG6}), (\ref{EQG7}) and since $(T,R)$ is a companion factorization of $X_z$, we have
$a^{r_{p_1,q}}za^{v_{p,q_1}} = a^{r_{p_2,q}}za^{v_{p, q_2}}$,
that is,
\begin{equation}
r_{p_1,q} = r_{p_2, q}, \quad v_{p,q_1} =  v_{p,q_2} \notag
\end{equation}
Thus the word
$$a^{r_{p_1,q}}za^{v_{p,q_1}} = a^{r_{p_1,q}}wa^{v_{p_1,q}}a^{r_{p,q_1}}wa^{v_{p,q_1}} =
a^{r_{p_2,q}}wa^{v_{p_2,q}} a^{\alpha n} a^{r_{p,q_2}}wa^{v_{p,q_2}} = a^{r_{p_2,q}}za^{v_{p, q_2}}$$
has two different factorizations in words of $X$. This is a contradiction since
$X$ is a code.
A similar argument applies to the sums $R_{p,(w)} + T_{q',(w')}$,
$R_{p',(w')} + T_{q,(w)}$ and the proof is complete.
\enddim

\begin{remark}
Notice that Eqs. (\ref{EQG1}), (\ref{EQG2}) are very similar to
Eq. (7) in \cite[Proposition 6.1] {CDF05}.
\end{remark}

%%%%%%%%%%%%%%%%%%%%%%%%%%%%%%%%%%%%%%%%%%%%%%%%%%%%%%%%%%%%%%%%%%%%%%%%%%

\section{The triangle conjecture} \label{TrCj}

The triangle conjecture has been originally stated as follows.

\begin{conjecture} [The Triangle conjecture]
Let $A =\{a, b \}$ a two-letter alphabet. Let $X$ be a finite subset
of $a^*ba^*$. Let $d = \max \{|x| ~|~ x \in X \}$.
If $X$ is a code, then $\Card(X) \leq d + 1$.
\end{conjecture}

It appears for the first time in \cite{PS77b} as a special case of the
commutative equivalence conjecture mentioned in Section \ref{IN}.
Indeed, in \cite{PS77b} the authors proved that a finite subset $X$
of $a^*ba^*$ is commutatively prefix if and only if
\begin{equation}
\forall k \geq 0 \quad \Card(\{a^rba^v \in X ~|~ 0 \leq r + v \leq k \}) \leq k + 1 \notag
\end{equation}
(see also \cite[Proposition 14.6.3]{BPR}).
The triangle conjecture has also been formulated in a graph theoretical setting in \cite{PS81}.
As far as we known, the term ``the triangle conjecture'' appears for the first time in \cite{PS}.
It originates in the following construction: if one represents every word of the form $a^rba^v$ by
a point $(r, v) \in \N^2$, the set $\{a^rba^v ~|~ 0 \leq r + v \leq k \}$ is represented by the triangle
$\{(r, v) \in \N^2 ~|~ 0 \leq r + v \leq k \}$.

As already said, in 1985 Shor found a counterexample to the triangle conjecture \cite{Shor}.
Other counterexamples may be found in \cite{Cordero19}. Thus this conjecture, along with the commutative
equivalence conjecture and the factorization conjecture were restricted
to the smaller class of finite maximal codes and its subsets.

Let $X$ be a finite maximal code on the alphabet $A$ and let $a \in A$ be a letter such that $a^n \in X$.
The following inequalities have been introduced in \cite{ZHSH} and generalize
the triangle conjecture.

\begin{equation} \label{EQZH3}
\forall k \geq 0 \quad \Card(\{a^rwa^v \in X_w ~|~ r + v \leq k \}) \leq k + 1
\end{equation}

The following is Corollary 3.3 in \cite{ZHSH}.

\begin{corollary} \label{CorTr2}
Let $X \subseteq A^+$ be a finite maximal code, let $n$ be the order of $a \in A$.
Let $B = A \setminus \{a \}$. If $n$ is a prime number,
then for any $w \in B(a^*B)^*$, $X_w$ satisfies Eqs.(\ref{EQZH3}).
\end{corollary}

%%%%%%%%%%%%%%%%%%%%%%%%%%%%%%%%%%%%%%%%%%%%%%%%%%%%%%%%%%%%%%%%%%%%%%%%%%%%%%%%%%%

\section{Haj\'{o}s factorizations and good arrangements} \label{HFRC}

In this section we recall the definitions of Haj\'{o}s factorizations and
of special arrangements of words, named good arrangements, with the main related results.
This material will be used in Section \ref{TC}.
We will also provide relationships between results proved for factorizing codes
and results demonstrated in Section \ref{main}.

\subsection{Haj\'{o}s factorizations}

Let $G$ be a finite abelian group.
We recall that a subset $S$
of $G$ is {\it periodic} if there exists $g$ in $G \setminus 0$
such that
$g+S=S$.
A factorization $(T,R)$ of $G$ is {\it periodic} if at least
one factor is periodic.
When we consider the particular case of the cyclic groups, one has the following
obvious characterization of the periodic factorizations, which we need
in the sequel \cite{Fuchs}.

\begin{lemma} \label{periodic}
A pair $(T,R)$ is a periodic factorization of $\Zc$ if and only
if there exists a divisor $p$ of $n$, $p \not = n$, and a factorization
$(T,S)$ of $\Z/p\Z$ such that $R$ is the direct sum of $S$ and
$\{0, 1, \ldots, (n/p-1)\}p$.
\end{lemma}

Let $(T,R)$ be a periodic factorization of $\Zc$. A
divisor $p$ of $n$, $p \not = n$, is a {\it period} of $R$ if
$R$ is the direct sum of $S$ and $\{0, 1, \ldots, (n/p-1)\}p$,
where $(T,S)$ is a factorization of $\Z/p\Z$.
Haj\'{o}s conjectured that any factorization of a finite abelian group
was periodic. This conjecture was false.
Thereafter, cyclic groups
were classed in good groups and
bad groups.
A group is {\it good} (or it has the {\it Haj\'{o}s property})
if {\it any} of its factorization
is periodic, otherwise it is {\it bad}.
The integer $n$ is said to be a {\it Haj\'{o}s number}
if the group $\Zc$ has the Haj\'{o}s property \cite{BPR}.
In a sequence of different papers,
good cyclic groups were characterized
\cite{Szabo}. They are
$\Z/p^{n}q\Z$, $\Z/p^{2}q^{2}\Z$,
$\Z/p^{2}qr\Z$, $\Z/pqrs\Z$ and their
subgroups, where
$p$, $q$, $r$, $s$ are different primes.

In \cite{HA50a}, Haj\'{o}s gave a method,
slightly corrected later by Sands
in \cite{SA}, for the construction of
a class
of periodic factorizations which
contains all factorizations of a good
group.
As done in
\cite{CDF96}, we report this method
for the cyclic group $\Zc$ of order $n$ (Definition
\ref{HCG}).
The corresponding
factorizations will be named
{\it Haj\'{o}s factorizations}.
The operation $\circ$ intervenes:
for subsets
$S= \{ s_{1}, \ldots , s_{q} \}$, $T$ of $\Zc$,
$S \circ T$ denotes the family of subsets of $\Zc$
having the form
$\{ s_{i} + t_{i} \mid i \in \{ 1, \ldots , q \} \}$, where
$\{ t_{1}, \ldots ,t_{q} \}$ is any multiset of elements of $T$
having the same cardinality as $S$.

\begin{definition} \label{HCG}
Let $R,T$ be subsets of $\N$.
$(R,T)$ is a Haj\'{o}s factorization of
$\Zc$ if and only if there are $s$ different integers
$k_i$ that form a chain of divisors of $n$:
\begin{eqnarray} \label{EH1}
k_{0} &=& 1 \mid k_{1} \mid k_{2} \mid \ldots \mid k_{s} =n,
\end{eqnarray}
such that:
\begin{eqnarray} \label{EH2}
a^{R} & \in &
((( \frac{a-1}{a-1} \circ \frac{a^{k_1}-1}{a-1}) \cdot
\frac{a^{k_{2}}-1}{a^{k_1}-1}) \circ \ldots \cdot \ldots
\frac{a^{n}-1}{a^{k_{s-1}}-1} ),
\end{eqnarray}
\begin{eqnarray} \label{EH3}
a^{T} & \in &
((( \frac{a-1}{a-1} \cdot \frac{a^{k_1}-1}{a-1}) \circ
\frac{a^{k_{2}}-1}{a^{k_1}-1}) \cdot \ldots \circ \ldots
\frac{a^{n}-1}{a^{k_{s-1}}-1} ).
\end{eqnarray}
Furthermore we have
$R,T \subseteq \{0, \ldots , n-1 \}$.
\end{definition}

Notice that Haj\'{o}s construction does not provide all periodic factorizations of cyclic groups.
That is, Haj\'{o}s factorizations form a class strictly included in the class of periodic factorizations
\cite[p. 243] {CDF96}.
Furthermore, Haj\'{o}s method applies to $\Zc$ even if $\Zc$ is a bad group.

As said in Section \ref{FattG},
Krasner factorizations are the simplest examples of
factorizations of $\Zc$.
Theorem \ref{HC}
is one of the
results which allow us to link
factorizing codes, Haj\'{o}s factorizations and Krasner factorizations.
It has been stated in \cite{CDF96} and
it has been generalized to the case of more than two factors in \cite{SZSA}.

\begin{theorem} \label{HC}
Let $(R,T)$ be subsets of
$\{0, \ldots , n-1 \}$. The
following conditions are equivalent:
\begin{itemize}
\item [1)]
$(R,T)$ is a Haj\'{o}s factorization of
$\Zc$.
\item [2)]
There exists
a Krasner factorization $(I,J)$ of
$\Zc$ such that $(I,T)$, $(R,J)$ are
(Haj\'{o}s) factorizations of $\Zc$.
\item [3)]
There exist
$L,M \subseteq \N$ and a
Krasner factorization $(I,J)$ of
$\Zc$
such that:
\begin{eqnarray} \label{EF}
a^R &=& a^{I}(1+a^{M}(a-1)), \quad
a^T=a^{J}(1+a^{L}(a-1)).
\end{eqnarray}
\end{itemize}
Furthermore,
$2) \Leftrightarrow 3)$ also
holds for $R,T \subseteq \N$.
\end{theorem}

As stated in Theorem
\ref{HC}, the equivalence between conditions
$2)$ and $3)$
still holds under the more general
hypothesis that $R,T$
are arbitrary subsets
of $\N$
(not necessarily with $\max R < n$,
$\max T < n$).
Then,
for $R,T \subseteq \N$,
we will say that $(R,T)$ is a Haj\'{o}s
factorization
of
$\Zc$ if
$(R_{(n)}, T_{(n)})$
satisfies the conditions contained in
Definition \ref{HCG}.
This is equivalent, as
Lemma 2.1 in \cite{CDF01} shows,
to defining
Haj\'{o}s
factorizations of
$\Zc$
as those pairs satisfying
Eqs.(\ref{EF}).

Theorem \ref{HC} points out that for each
Haj\'{o}s factorization $(R,T)$, we can associate
a Krasner factorization $(I,J)$ with
it, called a {\it Krasner companion
factorization} of
$(R,T)$ \cite{LAM97}.
This pair $(I,J)$ is uniquely defined by
the chain of divisors in Eq. (\ref{EH1}).
Indeed, let us consider the pair $(I,J)$ where $a^I$
is obtained by erasing
from Eq. (\ref{EH2}) polynomials
$P_j=(a^{k_{j}}-1)/(a^{k_{j-1}}-1)$
with $j$ odd, and $a^J$
is obtained by erasing
from Eq. (\ref{EH3})
polynomials $P_j$ with $j$ even, that is
\begin{eqnarray} \label{EK1}
a^{I} & = & \prod_{j \; {\rm even} \;, 1 \leq j \leq s}
\frac{(a^{k_{j}}-1)}{(a^{k_{j-1}}-1)}, \quad
a^{J}= \prod_{j \; {\rm odd} \;, 1 \leq j \leq s}
\frac{(a^{k_{j}}-1)}{(a^{k_{j-1}}-1)}
\end{eqnarray}
As proved in \cite{KRR}, $(I,J)$ is a Krasner factorization
of $\Zc$ and moreover, by Definition \ref{HCG},
$(I,T)$, $(R,J)$ are
(Haj\'{o}s) factorizations of $\Z_{n}$.
By Theorem \ref{HC},
$(R,T)$ satisfies Eqs.(\ref{EF}).
A proof by induction on
the length of the
chain in Eq. (\ref{EH1}) allows us to state that
$(I,J)$ is the unique such pair. In conclusion, we have the following result
which strengthens Proposition 4.2 in \cite{CDF05}.

\begin{proposition} \label{unicapair}
Let $(R,T)$ be a Haj\'{o}s factorization
of $\Z_{n}$. Assume that $(R_{(n)}, T_{(n)})$ is defined by
the chain of divisors in Eq. (\ref{EH1}).
Then, there is a unique Krasner factorization $(I,J)$
of $\Z_{n}$ associated with $(R,T)$ and the chain in Eq. (\ref{EH1})
such that $(R,T)$ satisfies Eqs.(\ref{EF}).
\end{proposition}

We need the following result which also provides additional information
on the relationship between factorizations and chains of
divisors.

\begin{proposition} \label{unichain}
Let $(I,J)$ be a Krasner factorization of $\Z_{n}$.
There exists a unique chain of different divisors of $n$
\begin{eqnarray*}
k_{0} &=& 1 \mid k_{1} \mid k_{2} \mid \ldots \mid k_{s} =n
\end{eqnarray*}
such that $(I,J)$ satisfies Eq. (\ref{EK1}).
\end{proposition}
\begdim
Let $(I, J)$ be as in the statement. If $s = 1$, we are done.
Otherwise, as proved in \cite{CDF89}, one has $J = \{0, 1, \ldots, k_1 - 1 \} + J'$,
where $k_1 > 1$,  $I, J'$ are divisible by $k_1$ and $k_1 \not \in J'$.
Moreover, $(I/k_1, J'/k_1)$ is a Krasner factorization of $\Z/(n/k_1)\Z$ defined by a shortest chain.
Of course, if there were another chain of divisors of $n$ defining $(I,J)$, the smallest divisor other than $1$
in this chain should be $k_1$. Then we conclude the proof by induction on $s$.
\enddim

There are at least two different recursive constructions
of the Haj\'{o}s factorizations.
In the next, we will use the one illustrated in Proposition \ref{HR}
\cite{LAM97,CDF06}.

\begin{proposition} \label{HR}
Let $R,T \subseteq \{0, \ldots , n-1 \}$
and suppose that $(R,T)$
is
a Haj\'{o}s factorization of
$\Zc$
with respect to the
chain
$k_{0} = 1 \mid k_{1} \mid k_{2} \mid \ldots \mid k_{s} =n$
of divisors of $n$.
Then either $(R,T) = (R_1, T_1)$
or $(R,T) = (T_1, R_1)$,
where $(R_1, T_1)$
satisfies one of the two following
conditions:
\begin{enumerate}
\item[1)]
There exists $r \in \{0, \ldots , n-1 \}$
such that
$R_1 =\{ r \}$ and
$T_1 =\{0, \ldots , n-1 \}$.
Furthermore, $s=1$.
\item[2)]
$R_1 \in R^{(1)} \circ \{0, 1, \ldots , g-1\}h$,
$T_1=T^{(1)} + \{0, 1, \ldots , g-1\}h$,
$(R^{(1)}, T^{(1)})$ being a Haj\'{o}s factorization
of $\Z/h\Z$, $g,h \in \N$,
$n=gh$,
$R^{(1)}, T^{(1)} \subseteq \{0, \ldots , h-1 \}$.
The chain of divisors defining
$(R^{(1)}, T^{(1)})$ is
$k_{0} = 1 \mid k_{1} \mid k_{2} \mid \ldots \mid k_{s-1}=h$.
\end{enumerate}
\end{proposition}

%%%%%%%%%%%%%%%%%%%%%%%%%%%%%%%%%%%%%%%%%%%%%%%%%%%%%%%%%%%%%%%
\subsection{Good arrangements} \label{definizioniGA}

Assume that a Krasner factorization $(J, I)$ of $\Zc$
is a companion factorization of a finite maximal code $X$ containing $a^n$, as in the case
of a factorizing code.
By Theorems \ref{endchinois}, \ref{HC}, for any $w \in B(a^*B)^*$,
there is an arrangement of the words of $X_w$ in a matrix such that
all pairs $(R_h,T_k)$, where $R_h$ is a row and $T_k$ is a column, are Haj\'{o}s factorizations of
$\Zc$. The integer $n$ is named here the {\it size} of the matrix.

In fact we can point out that this arrangement has additional properties, through the notion of
a {\it good arrangement} of a set.
This is a slightly technical notion introduced in \cite{CDF99,CDF05}.
Roughly, it is an arrangement of its words in a matrix
that can be recursively constructed thanks to the recursive constructions of the pairs
$(R_h,T_k)$.
The original definition has been given for subsets of $a^*(A \setminus \{a \})a^*$ but here
we extend it to subsets of $X_w$, where $X$ is a set containing $a^n$.
In \cite{CDF05}, it has been proved
that if $X$ is a factorizing code, then for every $b \in B$, $X_b$ has a good arrangement.
Reading the proofs in \cite{CDF05}, one can see that the existence of a good arrangement
depends only on the fact that a factorizing code has a companion factorization
which is a Krasner pair.
Hovewer, we give a new simpler proof of this result, namely that for a finite maximal
code $X$, the existence of a good arrangement of $X_w$
is equivalent to the existence of Krasner factorization
which is a companion factorization of $X$ (Proposition \ref{krcompanionGood}).

Let us introduce some notations used in Definition \ref{GAC}.
In this context, by abuse of notation, $X_w$ stand for any finite subset of
$a^*wa^*$, where $w \in B(a^*B)^*$.
Let $\mathcal{X}_w =
(a^{r_{p,q}}wa^{v_{p,q}})_{1 \leq p \leq m, \; 1 \leq q \leq \ell}$
be an arrangement of $X_w$. Then $\overline{\mathcal{X}_w} =
(a^{\overline{r_{p,q}}}wa^{\overline{v_{p,q}}})_{1 \leq p \leq m, \; 1 \leq q \leq \ell}$.
As classically known, the transpose matrix of $\mathcal{X}_w$ is the matrix obtained
by switching its word-rows with its word-columns.

For a nonnegative integer $h$, $a^hX_w$ denotes the matrix whose $pq$-component
is $a^{h + r_{p,q}}wa^{v_{p,q}}$.
If $\mathcal{Y}_w =
(a^{s_{p,q}}wa^{t_{p,q}})_{1 \leq p \leq m, \; 1 \leq q \leq \ell}$ is an arrangement
of $Y_w$, we set $\mathcal{X}_w \cup \mathcal{Y}_w$ the arrangement of
$X_w \cup Y_w$ having $m$ word-rows and $2 \ell$ columns,
and $pq'$-component equal to $a^{r_{p,q}}wa^{v_{p,q}}$
for $q' = 1, \ldots, \ell$, equal to $a^{s_{p,q'-\ell}}wa^{t_{p,q'-\ell}}$
for $q' = \ell+ 1, \ldots, 2 \ell$.

Then, given a matrix
$\mathcal{A} = (a_{p,q})_{1 \leq p \leq m, \; 1 \leq q \leq \ell}$
with entries
in $\N$
and an integer $n$, $n \geq 2$,
we set $\overline{\mathcal{A}} = (\overline{a_{p,q})}_{1 \leq p \leq m, \; 1 \leq q \leq \ell}$
and
$h + \mathcal{A} = (b_{p,q})_{1 \leq p \leq m, \; 1 \leq q \leq \ell}$,
where, for each $p, q$,
$1 \leq p \leq m, \; 1 \leq q \leq \ell$,
we have $b_{p,q} = h + a_{p,q}$.
The meaning of $\mathcal{A} \cup \mathcal{B}$, where
$\mathcal{B} = (a_{s,t})_{1 \leq s \leq m, \; 1 \leq t \leq \ell}$, is clear.

The following definitions are from \cite{CDF05}.

\begin{definition} \label{GA}
Let $(R_1,T_1), \ldots , (R_m,T_m)$
be Haj\'{o}s
factorizations
of
$\Zc$ having $(I,J)$
as a Krasner companion factorization.
An arrangement $\mathcal{D} = (r_{p,q})_{1 \leq p \leq m, \; 1 \leq q \leq
l}$
of $\cup_{p=1}^m R_p$
having the $R_p$'s as rows
is a {\rm good arrangement} of
$(R_1, \ldots , R_m)$ ({\rm with respect to the rows})
if $\mathcal{D}$ can be recursively constructed
by using the following three rules.
\begin{enumerate}
\item[1)]
$\mathcal{D}$ is a good arrangement of
$\cup_{p=1}^m R_p$ (with respect to the rows)
if so is
$\overline{\mathcal{D}}$
\item[2)]
Suppose that $(R_p,T_p)$ satisfies condition $1)$ in
Proposition \ref{HR}, for all $p \in \{1, \ldots , m \}$.
If $R_p = \{ r_p \}$ with $r_p \in \{0, \ldots , n-1 \}$,
then $\mathcal{D}$
is the matrix with only one column
having $r_p$ as the $p$th entry.
If $R_p  =\{r_{p,0}, \ldots , r_{p,n-1} \}$
with $r_{p,i} = i$, then
$\mathcal{D} =
(r_{p,j})_{1 \leq p \leq m, \; 0 \leq j \leq n-1}$.
\item[3)]
Suppose that $(R_p,T_p)$ satisfies
condition $2)$ in
Proposition \ref{HR}, for all $p \in \{1, \ldots , m \}$, i.e.,
either
$R_p=R^{(1)}_{p} + \{0, h, \ldots , (g-1)h)\}$
or
$R_p \in R^{(1)}_{p} \circ \{0, h, \ldots , (g-1)h)\}$.
Let $\mathcal{D}^{(1)}$ be a good arrangement of
$\cup_{p=1}^m R^{(1)}_p$. In the first case, we set
$\mathcal{D} = \cup_{k = 0}^{g-1} (kh + \mathcal{D}^{(1)})$.
In the second case,
$\mathcal{D}$
is obtained by taking $\mathcal{D}^{(1)}$
and then substituting in it each $r_{p,q}^{(1)} \in R_p^{(1)}$
with the corresponding $r_{p,q}^{(1)}+ \lambda_{p,q} h \in R_p$.
\end{enumerate}
\end{definition}

Let $(R_1,T_1), \ldots , (R_m,T_m)$
be Haj\'{o}s
factorizations
of
$\Zc$ having $(I,J)$
as a Krasner companion factorization.
Obviously, we can
consider arrangements of
$\cup_{p=1}^m R_p$
having the $R_p$'s as {\it columns} and
therefore,
we can
give a dual
notion of a {\it good arrangement} of
$\cup_{p=1}^m R_p$
{\it with respect to the columns}
(by using a corresponding dual operation
$\cup$).
This arrangement
will be
the transpose matrix of a good
arrangement of
$\cup_{p=1}^m R_p$ with respect to the rows.

\begin{definition} \label{GAC}
Let $w \in B(a^*B)^*$.
An arrangement
$\mathcal{X}_w =
(a^{r_{p,q}}wa^{v_{p,q}})_{1 \leq p \leq m, \; 1 \leq q \leq \ell}$
of a finite set $X_w \subseteq a^*wa^*$
is a {\rm good arrangement
(with $(I,J)$ as
a Krasner associated
pair)}
if it satisfies the
following three conditions:
\begin{enumerate}
\item[1)]
For each row $R_p$ and each column
$T_q$, $1 \leq p \leq m$, $1 \leq q \leq \ell$,
$(R_p,T_q)$ is a Haj\'{o}s factorization of
$\Zc$ having $(I,J)$ as a Krasner
companion factorization
with respect to a chain of divisors
of $n = \Card(X_w)$.
\item[2)]
The induced arrangement of the rows
is a good arrangement
of $\cup_{p=1}^m R_p$
with respect to the rows.
\item[3)]
The induced arrangement of the columns
is a good arrangement
of $\cup_{q=1}^l T_q$
with respect to the columns.
\end{enumerate}
\end{definition}

\begin{remark}
Note that condition 1) in Definition \ref{GAC}
is not sufficient to define a good arrangement. In
\cite[Example 7.2] {CDF05} there is an example of a set
that does not have good arrangements but that has arrangements
satisfying condition 1).
\end{remark}

The following result provides an easy example of a set that has a good arrangement.

\begin{proposition} \label{HajGood}
For any Haj\'{o}s factorization $(R, T)$ of
$\Zc$, for any $w \in B(a^*B)^*$,
the set $a^R w a^T$
has a good arrangement with $(I,J)$ as
associated
pair, where $(I,J)$ is a Krasner
companion factorization of $(R,T)$.
\end{proposition}
\begdim
By Definition \ref{GAC}, we may assume that $R,T \subseteq \{0, 1, \ldots, n-1 \}$.
The proof is by induction on the length of a chain of divisors of $n$ which defines
$(R, T)$.
If $(R,T)$ satisfies condition $1)$ in
Proposition \ref{HR}, a good arrangement of $a^R w a^T$
clearly exists.
Suppose that $(R,T)$ satisfies
condition $2)$ in
Proposition \ref{HR} and
$R = R^{(1)} + \{0, h, \ldots , (g-1)h)\}$,
$T \in T^{(1)} \circ \{0, h, \ldots , (g-1)h)\}$.
By induction hypothesis, there is a good arrangement of
$a^{R^{(1)}}w a^{T^{(1)}}$, hence the same holds
for $a^{R^{(1)}}w a^T$. Finally, if $\mathcal{Y}^{(1)}$
is a good arrangement of $a^{R^{(1)}}w a^T$, then
$\mathcal{Y} = \cup_{k = 0}^{g-1} (kh + \mathcal{Y}^{(1)})$
is a good arrangement of $a^R w a^T$.
If $R \in R^{(1)} \circ \{0, h, \ldots , (g-1)h)\}$,
$T = T^{(1)} + \{0, h, \ldots , (g-1)h)\}$, the above argument yields a good arrangement of
$a^T w a^R$ whose transpose is the required good arrangement of $a^R w a^T$.
\enddim

%%%%%%%%%%%%%%%%%%%%%%%%%%%%%%%%%%%%%%%%%%%%%%%%%%%%%%
\section{New results} \label{TC}

Let $X$ be a finite maximal code containing $a^n$.
We begin this section by showing that
the case where a companion factorization of $X$ is a Krasner factorization
is equivalent to the existence of a good arrangement of the words of $X_w$,
for any $w$ (Proposition \ref{krcompanionGood}).
Then, we consider the special case where $n$ is
the product of at most two (not necessarily distinct) prime numbers.
We prove that $X$ always has a companion factorization which is a
Krasner factorization of $\Zc$
(Proposition \ref{haj-kr2primes}). Hence, by Proposition
\ref{krcompanionGood}, every $X_w$ has a good arrangement.
Next, by using a result from \cite{CDF07b},
we state that a finite code
$Y \cup a^{pq} \subseteq a^*Ba^* \cup a^*$
is included in a factorizing code if and only if it is included in a finite maximal code,
and both conditions are equivalent to the existence of a good arrangement
for $Y$ (Proposition \ref{inclusion}).
Consequently we state that it is decidable whether
$Y \cup a^{pq}$ is included in a finite maximal code because the existence
of a good arrangement for $Y$ is a decidable property.

The following is Proposition 7.3 in \cite{CDF05}.

\begin{proposition} \label{krcompanion}
Let $X$ be a factorizing code containing $a^n$, let
$(J, I)$ be a Krasner factorization of $\Zc$ which
is a companion factorization of $X$.
Then, there is a good arrangement of $X_b$
with $(I,J)$ as a Krasner associated pair.
\end{proposition}

The following proposition generalizes the above result.

\begin{proposition} \label{krcompanionGood}
Let $X$ be a finite maximal code containing $a^n$.
There is a Krasner factorization $(J, I)$ of $\Zc$ which
is a companion factorization of $X$ if and only if each $X_w$ has a good arrangement
with $(I,J)$ as a Krasner associated pair, for any $w \in B(a^*B)^*$.
\end{proposition}
\begdim
Let $X$ be a finite maximal code containing $a^n$.
Looking at Definition \ref{GAC}, one can prove by induction that if each $X_w$ has a good arrangement
with $(I,J)$ as a Krasner associated pair, then $(J, I)$ is a companion factorization of $X$.
Conversely, assume that there is a Krasner factorization $(J, I)$ of $\Zc$ which
is a companion factorization of $X$, thus of each $X_w$.
By Theorem \ref{endchinois}, there is a bijection from $a^Iwa^J$
onto $X_w$. Moreover,
there is an arrangement
$\mathcal{X}_w = (a^{r_{p,q}}wa^{v_{p,q}})_{1 \leq p \leq m, \; 1 \leq q \leq \ell}$
of $X_w$ such that
for each row $R_p$ and and each column
$T_q$, the pairs
$(J , R_p)$, $(R_p,T_q)$, $(T_q, I)$ are all factorizations of
$\Zc$. Furthermore, by Theorem \ref{HC}, $(J , R_p)$, $(R_p,T_q)$, $(T_q, I)$ are all
Haj\'{o}s factorizations of $\Zc$ defined by the same chain of divisors of $n$, namely
the one uniquely determined by $(I,J)$ (Proposition \ref{unichain}).
In the proof of Theorem \ref{endchinois}, $\mathcal{X}_w$ is constructed starting from an arrangement of
$a^Iwa^J$. We know that $a^Iwa^J$ has a good arrangement $\mathcal{Y}$ (Proposition \ref{HajGood}).
It is easy to see that if $\mathcal{X}_w$ is constructed starting from $\mathcal{Y}$,
then $\mathcal{X}_w$ is a good arrangement of $X_w$ (all $R_p$ are periodic with the same period $h$ or
all $T_q$ are periodic with the same period $h$, depending on whether
$I$ is periodic with period $h$ or $J$ is periodic with period $h$).
\enddim

%%%%%%%%%%%%%%%%%%%%%%%%%%%%%%%%%%%%%%%%%%%%%%%%%%%%%%%%%%%%%
\subsection{Short chains of divisors} \label{particular}

Let $n \in \N$. In the following $\Omega(n)$ will be
the number of factors in the prime
factorization
of $n$. Recall that if $\Omega(n) \leq 2$, then $n$ is a
a Haj\'{o}s number and all the factorizations of $\Zc$ are
Haj\'{o}s factorizations.

\begin{proposition} \label{haj-kr2primes}
Let $X$ be a finite maximal code with $a^n \in X$.

If $\Omega(n) \leq 2$, then there is a factorization $(T,R)$
of $\Zc$ that is a companion
factorization of $X$ and, for any $w \in B(a^*B)^*$,
there is an arrangement
$\mathcal{X}_w = (a^{r_{h,k}}wa^{v_{h,k}})_{1 \leq h \leq m, \; 1 \leq k \leq \ell}$
of $X_w$ such that for each row $R_h = \{r_{h,k} ~|~ k \in \{1, \ldots , \ell \} \}$
and each column
$T_k =\{v_{h,k} ~|~ h \in \{1, \ldots , m \} \}$ one has
\begin{itemize}
\item[(1)]
the pairs
$(T , R_h)$, $(R_h,T_k)$, $(T_k, R)$ are Haj\'{o}s factorizations of
$\Zc$ defined by the same chain of divisors of $n$, hence they have
a same Krasner companion factorization $(I, J)$;
\item[(2)]
the pair $(I, J)$ is a companion factorization of $X$.
\end{itemize}
\end{proposition}
\begdim
Let $X$ be a finite maximal code with $a^n \in X$. If
$(T,R)$ is a factorization of $\Zc$ that is a companion
factorization of $X$, then
by Theorem \ref{endchinois}, for any $w \in B(a^*B)^*$,
there is an arrangement
$\mathcal{X}_w = (a^{r_{h,k}}wa^{v_{h,k}})_{1 \leq h \leq m, \; 1 \leq k \leq \ell}$
of $X_w$ such that for each row $R_h = \{r_{h,k} ~|~ k \in \{1, \ldots , \ell \} \}$
and each column
$T_k =\{v_{h,k} ~|~ h \in \{1, \ldots , m \} \}$
the pairs
$(T , R_h)$, $(R_h,T_k)$, $(T_k, R)$ are factorizations of
$\Zc$. If $\Omega(n) \leq 2$, these pairs
are Haj\'{o}s factorizations of
$\Zc$.

We may assume $T, R \subseteq \{ 0, \ldots , n-1 \}$.
Indeed, if $(T,R)$ is a Haj\'{o}s factorization of $\Zc$, then so is
$(T_{(n)}, R_{(n)})$. Moreover, if $(T,R)$
is a companion factorization of $X$, then $(T_{(n)}, R_{(n)})$
is also a companion factorization of $X$.

Suppose that $(T,R)$ is defined by a chain of divisors of $n$ of length one.
By Proposition \ref{HR}, $T = \{t \}$, with $t \in \{0, \ldots , n-1 \}$ and
$R =\{0, \ldots , n-1 \}$ (the other case is symmetric).
Since $(T,R)$
is a companion factorization of $X$, it is clear that one has
\begin{equation}
\{a^{\overline{i}}wa^{\overline{j}} ~|~ a^iwa^j \in X_w \} =
\{a^iwa^{v_i} ~|~ v_i \in \{0, \ldots , n-1 \}, \; 0 \leq i \leq n -1 \} \notag
\end{equation}
Thus, $\mathcal{X}_w$ has a unique row $R_1$ such that $R_1 = R \pmod{n}$. Hence, for each column $T_i = \{v_i \}$,
$0 \leq i \leq n -1$, the pairs
$(T , R_1)$, $(R_1,T_i)$, $(T_i, R)$ are Haj\'{o}s factorizations of
$\Zc$ defined by the same chain of divisors of $n$ and they all have
the Krasner pair $(\{ 0 \}, \{ 0, \ldots , n-1 \})$ as companion Krasner factorization.
This Krasner pair is clearly a companion factorization of $X_w$, for any $w$, hence it
is a companion factorization of $X$.

Suppose that $(T,R)$ is defined by a chain of divisors of $n$ of length greater than one,
thus $n = pq$ with $p,q$ not necessarily distinct prime numbers,
say $1 \mid p \mid pq$ (the other case is symmetric).
Hence
\begin{equation} \label{EqEq}
1 < \Card(R_h) = \Card(R) < n, \quad 1 < \Card(T) = \Card(T_k) < n
\end{equation}

We know that $(T , R_h)$, $(R_h,T_k)$, $(T_k, R)$ are
all periodic factorizations of $\Zc$.
We may assume that one of the following two cases occurs.
\begin{itemize}
\item[(a)]
$T$ is not periodic.
\item[(b)]
The sets $T,R_h ,T_k, R$ are all periodic.
\end{itemize}
Indeed, if $R$ is not periodic (and $T$ is periodic), then $(R, T)$ is a companion factorization
of $\widetilde{X}$ such that case (a) holds.
Of course, the proof that (1) and (2) hold for $\widetilde{X}$ also works to prove that
(1) and (2) hold for $X$.
Moreover, if $T, R$ are both periodic and there is $T_k$ (resp. $R_h$) which is not periodic,
then, by Theorem \ref{endchinois}, $(T_k, R)$ (resp. $(R_h, T)$) is a companion factorization of $X$
(resp. of $\widetilde{X}$) and we are in case (a).

Then in both cases (a) and (b), $R$ and all $R_h$'s are periodic.
Furthermore, it follows easily by Eq. (\ref{EqEq}) and Lemma \ref{periodic}
that $R$ and all $R_h$'s have the same period, say $p$.
Hence, one has
$R = \{ r \} + \{0, 1, \ldots , q-1\}p$, $R_h = \{ r_h \} + \{0, 1, \ldots , q-1\}p$,
$1 \leq h \leq m$, where
$r, r_h \in \{0, \ldots , p -1 \}$, $T, T_k \in \{0, \ldots , p- 1 \} \circ \{0, 1, \ldots , q-1\}p$,
$1 \leq k \leq \ell$.
Thus, by Definition \ref{HCG}, $(T , R_h)$, $(R_h,T_k)$, $(T_k, R)$ are all defined by a same chain of
divisors of $n$ and they all have
the Krasner pair $(\{0, \ldots , p- 1 \}, \{0, 1, \ldots , q-1\}p)$ as companion Krasner factorization.
This Krasner pair is clearly a companion factorization of $X_w$, for any $w$, hence it
is a companion factorization of $X$.
\enddim

%%%%%%%%%%%%%%%%%%%%%%%%%%%%%%%%%%%%%%%%%%%%%%%%%%%%%%%%%%%%%%%%%%%%
\subsection{A partial result on the inclusion problem} \label{IP}

The following is part of Corollary 7.1 in \cite{CDF07b}.

\begin{proposition} \label{GAandIncl}
Let $A = \{a, b_1, \ldots , b_m \}$, $m \geq 1$, an alphabet, let $B =  \{b_1, \ldots , b_m \}$.
Let $Y$ be a subset of $a^*Ba^*$ such that, for all ${\ell} \in \{1, \ldots , m\}$,
$\Card(Y \cap a^* b_\ell a^*) = n$ with
$\Omega(n) \leq 2$.
Then,
the following conditions are equivalent.
\begin{itemize}
\item[(1)]
There exists
a factorizing code $X \subseteq A^*$ such that
$Y = X \cap a^*Ba^*$.
\item[(2)]
There exists a Krasner factorization
$(I,J)$ of
$\Zc$
such that
$Y \cap a^*b_\ell a^*$ has a good arrangement
with $(I,J)$ as a Krasner associated
pair, for each ${\ell} \in \{1, \ldots , m\}$.
\end{itemize}
\end{proposition}

\begin{remark}
Proposition \ref{GAandIncl} is part of
Corollary 7.1 in \cite{CDF07b}.
What is missing is a third condition, equivalent to (1) and (2), which describes the structure of the factorizing code
$X$. Indeed, the proof constructs $X$.
Moreover, with the same notations as in Proposition \ref{GAandIncl}, we observe that if
$\Omega(n) \leq 2$ and $\Card(Y \cap a^* b_\ell a^*) = n$, then the existence of a good arrangement for
$Y$ ensures that $Y$ is code. This is not true for every $n$, a counterexample is
described in \cite[Proposition 4.1] {CDF07b}.
\end{remark}

The results demonstrated so far have an important consequence set out below.

\begin{proposition} \label{inclusion}
Let $A = \{a, b_1, \ldots , b_m \}$, $m \geq 1$, an alphabet, let $B =  \{b_1, \ldots , b_m \}$.
Let $Y$ be a subset of $a^*Ba^*$ such that, for all ${\ell} \in \{1, \ldots , m\}$,
$\Card(Y \cap a^* b_\ell a^*) = n$ with
$\Omega(n) \leq 2$.
Then,
the following conditions are equivalent.
\begin{itemize}
\item[1)]
There exists
a finite maximal code $X \subseteq A^*$ such that
$Y = X \cap a^*Ba^*$.
\item[2)]
There exists a Krasner factorization
$(I,J)$ of
$\Zc$
such that
$Y \cap a^* b_\ell a^*$ has a good arrangement
with $(I,J)$ as a Krasner associated
pair, for each ${\ell} \in \{1, \ldots , m\}$.
\item[3)]
There exists
a factorizing code $X \subseteq A^*$ such that
$Y = X \cap a^*Ba^*$.
\end{itemize}
Consequently, it is decidable whether $Y \cup a^n$ is included in a finite maximal code.
\end{proposition}
\begdim
Assume that $X \subseteq A^*$ is a finite maximal code such that
$Y = X \cap a^*Ba^*$ and, for all ${\ell} \in \{1, \ldots , m\}$,
$\Card(Y \cap a^* b_\ell a^*) = n$.
By Proposition \ref{nbaionette} the order of $a$ relative to $X$ cannot
be less than $n$ and by Proposition \ref{nelementi} it cannot be greater than $n$,
that is, $a^n \in X$. Thus, by
Propositions \ref{krcompanionGood}, \ref{haj-kr2primes},
1) implies 2). By Proposition \ref{GAandIncl}, 2) implies 3).
Of course, 3) implies 1).
Finally, by Proposition \ref{nelementi}, $Y \cup a^n$ is included in a finite maximal code
if and only if condition 1) holds and it is decidable whether $Y$ satisfies condition 2).
\enddim

%%%%%%%%%%%%%%%%%%%%%%%%%%%%%%%%%%%%%%%%%%%%%%%%%%%%%%%%%%%%%%%%%%%%%%%
\subsection{A partial result on the triangle conjecture}

Let $(I, J)$ be a Krasner factorization of $\Zc$.
Proposition \ref{Krasnertrue}
can be easily proved by induction or by observing that
$a^Iba^J \cup \{ a^n \}$
is a factorizing code and for any $w \in B(a^*B)^*$,
$a^Iwa^J \cup \{ a^n \}$
is the composition of two factorizing codes (see Proposition 14.1.2 in \cite{BPR}).

\begin{proposition} \label{Krasnertrue}
Let $(I, J)$ be a Krasner factorization of $\Zc$.
Then for any $w \in B(a^*B)^*$,
Eqs.(\ref{EQZH3}) hold for $a^Iwa^J$.
\end{proposition}

The proof of Proposition \ref{ZH3} is reported for the sake of completeness.
It is part of the proof of Theorem \ref{ZHSHDiscreteMath} \cite[Theorem 3.1] {ZHSH}.

\begin{proposition} \label{ZH3}
Let $A = B \cup \{a \}$ be an alphabet.
Let $w \in B(a^*B)^*$ and let $X_w \subseteq a^*ba^*$.
Assume that there are a Krasner factorization
$(I,J)$ of $\Zc$ and a bijection $\phi : X_w \rightarrow a^Iwa^J$ such that, for each
$a^rwa^v \in X_w$,
if $\phi(a^rwa^v) = a^{i_r}wa^{j_v}$,
then $r \geq i_r$, $v \geq j_v$.
Then $X_w$ satisfies Eqs.(\ref{EQZH3}).
\end{proposition}
\begdim
Let $I, J, X_w, \phi$ be as in the statement.
Then one has
\begin{eqnarray*}
\Card(\{a^rwa^v \in X_w ~|~ r + v \leq k \}) & = &
\Card(\{ a^{i_r}wa^{j_v} \in \phi(X_w) ~|~ r + v \leq k \}) \\
&& \mbox{ (because $\phi$ is a bijection)} \\
& = & \Card(\{ a^{i_r}wa^{j_v} \in a^Iwa^J ~|~ r + v \leq k \}) \\
&& \mbox{ (because } \phi(X_w) =  a^Iwa^J) \\
& \leq &
\Card(\{ a^{i}wa^{j} \in a^Iwa^J ~|~ i + j \leq k \}) \\
&& \mbox{ (because } r \geq i_r, v \geq j_v ) \\
& \leq & k + 1 \mbox{ (by Proposition \ref{Krasnertrue}) }
\end{eqnarray*}
\enddim

\begin{proposition} \label{triangleKrasner}
Let $X \subseteq A^*$ be a finite maximal code with $a^n \in X$.
Let $B = A \setminus \{a \}$.
If there is a Krasner factorization
$(I,J)$ of $\Zc$ which is
a companion factorization of $X$,
then for any $w \in B(a^*B)^*$, $X_w$ satisfies Eqs.(\ref{EQZH3}).
\end{proposition}
\begdim
Let $X, n, B, I, J$ be as in the statement. By Proposition \ref{ZH3}
it suffices to prove that there is a bijection $\phi : X_w \rightarrow a^Iwa^J$ such that, for each
$a^rwa^v \in X_w$,
if $\phi(a^rwa^v) = a^{i_r}wa^{j_v}$,
then $r \geq i_r$, $v \geq j_v$. In order to do that, we may assume $X_w = \overline{X_w}$, where
$\overline{X_w} = \{a^{\overline{r}}wa^{\overline{v}} ~|~ a^rwa^v \in X_w \}$
By Theorem \ref{endchinois}, there is a bijection from $a^Iwa^J$
onto $X_w$.
As in the proof of Proposition \ref{krcompanionGood}, we define this bijection $\phi$
and the corresponding (good) arrangement $\mathcal{X}_w$ of $X_w$
starting from a good arrangement of $a^Iwa^J$.
A technical proof by induction based on the definition of a good arrangement shows that
$\phi$ has the required property.
\enddim

Corollary \ref{triangle2primes} is a direct consequence
of Propositions \ref{haj-kr2primes} and \ref{triangleKrasner}.

\begin{corollary} \label{triangle2primes}
Let $X$ be a finite maximal code with $a^n \in X$ and $\Omega(n) \leq 2$.
Then for any $w \in B(a^*B)^*$, $X_w$ satisfies Eqs.(\ref{EQZH3}).
\end{corollary}

%%%%%%%%%%%%%%%%%%%%%%%%%%%%%%%%%%%%%%%%%%%%%%%%%%%%%%%%%%%%%%%%%%%%%%%%%%%%%%%%%%%%%%%%%%%%%%%%%%

\section{Future Perspectives} \label{end}

In this section we summarize the main issues arising from this research.
Obviously the aim is the solution of the three conjectures presented in Section \ref{IN}.
A preliminary step in this direction could be to extend some results of this paper.

First, in Proposition \ref{haj-kr2primes}, we state that $X$ always has a companion factorization which is a
Krasner factorization of $\Zc$ when the order $n$ of $a$ is a product
of at most two prime numbers.
This result guarantees
that the triangle conjecture (``basic'' or stronger version) is true for $X_w$.
We wondered if Proposition \ref{haj-kr2primes} can be extended to the case where
$X$ has a companion factorization which is a Haj\'{o}s factorization of $\Zc$, or, at least,
when $n$ is a Haj\'{o}s number.

Next, once again if $n$ is a prime number or
a product of two (not necessarily distinct) prime numbers, by using
Corollary 7.1 in \cite{CDF07b},
we state that a finite code
$Y \cup a^{n} \subseteq a^*Ba^* \cup a^*$
is included in a factorizing code if and only if it is included in a finite maximal code.
The proof of Corollary 7.1 in \cite{CDF07b} provides a construction
of this factorizing code, if it exists.
One might ask if this construction could be extended to provide all the finite maximal codes
in this small family, namely that of finite maximal codes such that the order of $a$ is a product
of at most two prime numbers.

Then, all partial results on the triangle conjecture require that the factorizations
of cyclic groups associated with $X_w$ are Haj\'{o}s factorizations.
Is there a finite maximal code for which this does not happen?
If there were, it would be a counterexample to the factorizing conjecture.

Finally, in Section \ref{main} we have given a proof of Theorem \ref{ZHmain} different from that in
\cite{ZHSH}.
In both proofs there are pairs of sets of integers
defined starting from strongly right completable words (with respect to a finite maximal code).
It would be interesting to investigate whether there is a relationship between
the pairs of sets $(P,Q)$ defined in \cite{ZHSH} and those $(T, R)$ defined here.

%%%%%%%%%%%%%%%%%%%%%%%%%%%%%%%%%%%%%%%%%%%%%%%%%%%%

\section{Appendix} \label{App}

It is noteworthy to point out that
Theorem 2.9 in \cite{ZHSH} gives more, as detailed below.
The following notions have been introduced in \cite{ZHSH}.
\begin{definition}
The set $a^P$ is a {\rm left set} of $X$ if there is a strongly right completable word
$y \in A^*$ for $X$ such that
$$P = \{ i \in T ~|~ ya^{2n|X| + i} \in X^* \}.$$
The set $a^Q$ is a {\rm right set} of $X$ if there is a word $x \in A^*$ such that
\begin{eqnarray*}
Q &=& \{ k \in T ~|~ a^{k +2n|X|}xA^* \cap X^* \not = \emptyset \} \quad \mbox{and} \\
\forall i,j \in Q, i < j && a^{j-i} \not \in (X^*)^{-1}X^*
\end{eqnarray*}
The word $y$ (resp. $x$) is the generator of the left (resp. right) set $a^P$ (resp. $a^Q$).
\end{definition}
Let $X \subseteq A^*$ be a finite maximal code and let $n$ be the order
of $a \in A$.
In \cite{ZHSH}, the authors proved that, for any left set $a^P$ of $X$ and any right set $a^Q$ of $X$,
the pair $(P,Q)$ is a factorization of $\Zc$.
Furthermore, it is one of these pairs that L. Zhang and P. K. Shum have
used in the proof of Theorem \ref{ZHmain}.

We proved Theorem \ref{ZHmain} by using the pair $(R, T)$ defined
in Proposition \ref{fattcompagna}. Of course the number of these pairs is related to the number
of factorizations of $X$ as in Theorem \ref{WFc}. In turn, these factorizations are also related
to strongly right completable words with respect to $X$, as pointed out in Remark \ref{vuota}.
Therefore, a question that naturally arises is
whether there is a relationship between these pairs $(R, T)$ and those
$(P,Q)$ formed by a left and a right set.

Finally there is an additional item in the statement of Theorem \ref{ZHmain} that we report below.
{\it
\begin{itemize}
\item[(4)]
For any $P$ in $\{P_1, \ldots , P_t \}$, $a^{P_{(n)}}$ is a left set of $X$,
and for each $Q$ in $\{Q_1, \ldots , Q_s \}$, $a^{Q_{(n)}}$ is a right set of $X$.
\end{itemize}
}

The
following notions that have been introduced in \cite{ZHSH}.

\begin{definition} \label{D1}
Let $\mathbb{P} \subseteq \mathcal{P}(\N)$ and $\mathbb{Q} \subseteq \mathcal{P}(\N)$,
where $\mathcal{P}(\N)$ is the powerset of $\N$. Let $n \in \N$.
The pair $(\mathbb{P},\mathbb{Q})$ is a {\rm system of factorizations} of $\Zc$
if for any $P \in \mathbb{P}$ and any $Q \in \mathbb{Q}$, $(P,Q)$
is a factorization of $\Zc$.
\end{definition}

\begin{definition} \label{D2}
Let $X$ be a finite maximal code, let $a$ be a letter of order $n$.
The {\rm system of factorizations} of $\Zc$
{\rm induced by} $X$ is the pair $(\mathbb{P} , \mathbb{Q})$, where
$$\mathbb{P} = \{ P \subseteq T ~|~ a^P \mbox{is a left set of } X \},
\quad \mathbb{Q} = \{ Q \subseteq T ~|~ a^Q \mbox{ is a right set of } X \}$$
\end{definition}

The following is Theorem 3.1 in \cite{ZHSH}. It is a partial result on the triangle conjecture.
It makes Definition \ref{D2} intervene.

\begin{theorem} \label{ZHSHDiscreteMath}
Let $X \subseteq A^+$ be a finite maximal code, let $n$ be the order of $a \in A$.
Let $B = A \setminus \{a \}$.
If the system of factorizations of $\Zc$ induced by $X$ contains a Krasner factorization
of $\Zc$, then for any $w \in B(a^*B)^*$, $X_w$ satisfies Eqs.(\ref{EQZH3}).
\end{theorem}

The following is Corollary 3.2 in \cite{ZHSH}.
It is a direct consequence of Theorem \ref{ZHSHDiscreteMath}.

\begin{corollary} \label{CorTr1}
Let $X \subseteq A^+$ be a finite maximal code, let $n$ be the order of $a \in A$.
Let $B = A \setminus \{a \}$.
If there exists a factorization $(P,Q)$ of $\Zc$ induced by $X$ such that one of $P$ and $Q$ is
a singleton, then for any $w \in B(a^*B)^*$, $X_w$ satisfies Eqs.(\ref{EQZH3}).
\end{corollary}

Theorem \ref{ZHSHDiscreteMath} was subsequently improved by the same authors. Precisely,
the following result is demonstrated in a submitted but unpublished article \cite[Theorem 3.1] {ZHSHResGat}.
Now Definitions \ref{D1}, \ref{D2} intervene.

\begin{theorem} \label{ZHSHResearchGate}
Let $X \subseteq A^+$ be a finite maximal code, let $n$ be the order of $a \in A$.
Let $B = A \setminus \{a \}$.
Let $(\mathbb{P} , \mathbb{Q})$ be the system of factorizations of $\Zc$
induced by $X$. If there exists $P \subseteq \N$ (resp. $Q \subseteq \N$) such that
$(\{P \} \cup \mathbb{P} , \mathbb{Q})$ (resp. $(\mathbb{P} , \{ Q \} \cup \mathbb{Q})$
is still a system of factorizations of $\Zc$ and it contains a Krasner factorizations of
$\Zc$, then for any $w \in B(a^*B)^*$, $X_w$ satisfies
Eqs.(\ref{EQZH3}).
\end{theorem}

%%%%%%%%%%%%%%%%%%%%%%%%%%%%%%%%%%%%%%%%%%%%%%%%%%%%%%%%%%%%
\bibliographystyle{plain}
\bibliography{CodesFactorization2021Arxiv}

\end{document}